\documentclass[12pt]{article}
\usepackage{graphicx}
\usepackage{amsmath, amsthm, amssymb}
\usepackage{multirow}
\usepackage{natbib, hyperref}
\usepackage{bm}
\usepackage{color, url, pdfpages}
\usepackage{geometry, setspace}
\newtheorem{assumption}{Assumption}
\allowdisplaybreaks[4]

\DeclareMathOperator\avar{avar}

\newtheorem{example}{Example}

\begin{document}

\title{\bf Inference for cumulative incidences and treatment effects in randomized controlled trials with time-to-event outcomes under ICH E9 (R1)}

\author{Yuhao Deng$^1$, Shasha Han$^2$ and Xiao-Hua Zhou$^3$* \\
{\small 1 University of Michigan \ 2 Chinese Academy of Medical Sciences \& Peking Union Medical College} \\ {\small 3 Peking University}}

\maketitle

\begin{abstract}
In randomized controlled trials (RCTs) that focus on time-to-event outcomes, intercurrent events can arise in two ways: as semi-competing events, which modify the hazard of the primary outcome events, or as competing events, which make the definition of the primary outcome events unclear. Although five strategies have been proposed in the ICH E9 (R1) addendum to address intercurrent events in RCTs, these strategies are not easily applicable to time-to-event outcomes when aiming for causal interpretations. In this study, we show how to define, estimate, and make inferences concerning objectives that have causal interpretations within these contexts. Specifically, we derive the mathematical formulations of the causal estimands corresponding to the five strategies and clarify the data structure needed to identify these causal estimands. Furthermore, we introduce nonparametric methods for estimating and making inferences about these causal estimands, including the asymptotic variance of estimators and hypothesis tests. Finally, we illustrate our methods using data from the LEADER Trial, which aims to investigate the effect of liraglutide on cardiovascular outcomes. \par
\end{abstract}

\section{Introduction} \label{sec1}

Intercurrent events refer to the events ``occurring after treatment initiation [of clinical trials] that affect either the interpretation of or the existence of the measurements associated with the clinical question of interest" \citep{ICH19}. In randomized controlled trials (RCT) with time-to-event outcomes, intercurrent events occur as either competing events, which prevent the primary outcome events, or semi-competing events, which modify the hazard of the primary outcome events. Five strategies have been proposed in the International Conference on Harmonization (ICH) E9 (R1) addendum to address intercurrent events, namely, treatment policy strategy, composite variable strategy, while on treatment strategy, hypothetical strategy, and principal stratum strategy. To answer a specific scientific question, a strategy with a particular estimand is chosen before the study design. For example, the treatment policy strategy investigates the effect of a ``treatment policy'', initial treatment plus the intercurrent events as natural. In contrast, the composite variable strategy investigates the treatment effect on a composite outcome, a variable determined jointly by the primary outcome event and intercurrent event \citep{rufibach2019treatment, kahan2024estimands, siegel2024time, fierenz2024current}.


Several studies have studied the five strategies for binary and continuous outcomes under the potential outcomes framework of causal inference \citep{lipkovich2020causal, ratitch2020choosing, stensrud2022translating, ionan2023clinical}. In particular, a recent study considered the finite sample estimands and elaborated on how to construct these estimands when possibly more than one intercurrent event is anticipated \citep{han2023defining}. However, the insights derived from binary and continuous outcomes do not readily extend to the context of time-to-event outcomes. In the context of binary and continuous outcomes, the outcomes are measured at a fixed time point within the study period, and the potential outcomes under the assigned treatment are observable at the individual level. The causal estimand is a direct contrast of potential outcomes, and it is time-independent. This investigation of causal effects can be extended to repeated measurements \citep{young2020causal, janvin2024causal}. However, due to censoring by loss of follow-up, the potential failure times under the assigned treatments may be unidentifiable when the outcomes are time-to-event, a problem especially relevant to individuals with long potential failure times. As such, the outcome variables of scientific interest often turn to the distribution of potential failure times, such as the cumulative incidence function or survival probability. Therefore, the causal estimands, which reflect the pointwise difference in the distributions of failure times between the treated and control groups, are time-dependent.

The time-dependent causal estimands bring challenges in estimation and inference. Existing literature in survival analysis has devoted considerable efforts to estimating survival functions when the outcomes are well-defined but possibly censored \citep{kaplan1958nonparametric, nelson1972theory, aalen1978nonparametric, cox1972regression, wei1992accelerated, lin1998additive, winnett2002adjusted}. Although it is routine to adjust for events that prevent the outcome measure, conventional techniques are often ad hoc and data-driven, and the resulting estimates may not have causal interpretations. For example, intercurrent events may be viewed as censoring events under the assumption that ``censoring'' is non-informative. This notion is inappropriate because the occurrence of intercurrent events may be related to potential outcomes of interest, rendering the non-informative ``censoring'' assumption implausible \citep{coemans2022bias}. Another way is to select a subsample with no intercurrent events. Since the risks of intercurrent events can vary across different groups, sample selection results in unbalanced underlying features between treatment groups, also known as selection bias \citep{stuart2011use, rudolph2014estimating}. The occurrence of intercurrent events may indicate treatment failure or rescue mediation, so causally interpreting the cumulative incidence of the primary outcome events must take the intercurrent events into account.

Moreover, traditional survival analysis has not widely integrated intercurrent events into causal estimands, unless when considering a pair of competing events, which includes both competing risks and semi-competing risks. The former views the intercurrent event and primary outcome event as two mutually exclusive events where the occurrence of one event prevents the occurrence of the other \citep{prentice1978analysis, gray1988class, fine1999proportional, andersen2012competing, austin2016introduction}. The latter considers one-sided exclusion, where the primary outcome event is a terminal event, and the intercurrent event is an intermediate event. The occurrence of the intercurrent event does not prevent the primary outcome event \citep{fine2001semi, haneuse2016semi, huang2021causal}. Some literature reviewed estimands in the presence of competing risks, such as marginal cumulative incidence, cause-specific cumulative incidence (subdistriution function) on the risk scale, as well as marginal hazard, subdistribution hazard, cause-specific hazard on the hazard scale \citep{gooley1999estimation, pintilie2007analysing, lau2009competing, geskus2015data, emura2020comparison, young2020causal}. In general, estimands on the risk scale lead to better causal interpretation than those on the hazard scale because timewise hazards are comparing individuals with different underlying features \citep{martinussen2013collapsibility, aalen2015does}. Due to the collapsibility of cumulative incidence, inference for the average treatment effect can be easily extended from randomized controlled trials to stratified randomized trials. Estimation methods to derive estimates were proposed under the frameworks of competing risks and semi-competing risks, including copula modelling, frailty modelling, maximum likelihood estimation, influence function based estimation and Bayesian inference \citep{peng2007regression, xu2010statistical, chen2010semiparametric, chen2012maximum, mao2017efficient, xu2022bayesian, nevo2022causal, gao2023defining, wei2023bivariate, martinussen2023estimation, yu2023unified, comment2019survivor}.

However, these works were more motivated by specific problems in analysis with (semi-)competing risks rather than providing a comprehensive understanding of the five strategies proposed by ICH E9 (R1) with time-to-event outcomes. Indeed, the medical community is calling for a thorough and rigorous approach to following the ICH E9 (R1) in the context of time-to-event data \citep{kahan2020treatment, cro2022evaluating}. It is essential to guide clinical practitioners in defining, estimating, inferring, and interpreting estimands under the strategies from a causal perspective. In this paper, we examine the scenario where a time-to-event primary outcome event and a potential time-to-event intercurrent event coexist. We describe the time-dependent causal estimands under the five strategies in ICH E9 (R1) for time-to-event outcomes. We provide nonparametric estimation, inference, and hypothesis testing methods in RCT with random censoring by adopting methodologies in survival analysis. Finally, we illustrate these five strategies using a real-world clinical trial, the LEADER (Liraglutide Effect and Action in Diabetes: Evaluation of cardiovascular outcome Results) Trial \citep{marso2016liraglutide}. The proposed estimation methods are easily implementable and are applicable to studying various scientific questions of interest. The insights of our proposed framework, which involves intervening on cause-specific hazards, can be easily extended to accommodate multiple types of intercurrent events by combining strategies for each intercurrent event. Studies of multiple endpoints and combining strategies to address multiple intercurrent events can help practitioners understand treatment effects more comprehensively.

The remainder of this paper is organized as follows. In Section \ref{sec2}, we define estimands under the strategies in ICH E9 (R1). In Section \ref{sec3}, we provide nonparametric Nelson--Aalen estimators and asymptotic variances to construct pointwise confidence intervals. In Section \ref{sec4}, we derive hypothesis testing methods based on log-rank tests. We summarize a comparison between these five strategies in Section \ref{sec5}, and apply these methods to the LEADER Trial in Section \ref{sec6}. Finally, we conclude this study with a brief discussion in Section \ref{sec7}. In the Supplementary Material, we extend the frameworks to observational studies by adjusting for baseline covariates and provide efficient influence function-based estimators with pointwise confidence intervals.

\section{Estimands for time-to-event outcomes} \label{sec2}

In this section, we describe the time-dependent causal estimands under the five strategies to address intercurrent events. We shall adopt the potential outcomes framework \citep{rubin1974estimating} that defines a causal estimand as the contrast between functionals of potential outcomes.
Consider a randomized controlled trial with $n$ individuals randomly assigned to one of two treatment conditions, denoted by $w$, where $w = 1$ represents the active treatment (a test drug) and $w = 0$ represents the control (placebo). An upper limit for the study duration of interest since the treatment initiation is set to be $t^*$. Assume that all patients adhere to their treatment assignments and do not discontinue treatment before $t^*$. For illustration, we assume there is at most one intercurrent event. Otherwise, we focus on the first intercurrent event. As such, associated with individual $i = 1, \ldots, n$ are two potential time-to-event primary outcomes $T_i(1)$ and $T_i(0)$, if any, which represent the time durations from treatment initiation to the primary outcome event under two treatment assignments respectively. Let $R_i(1)$ and $R_i(0)$ denote the occurrence time of potential intercurrent events, if any, under the two treatment assignments, respectively. 
If primary outcome events are prevented by intercurrent events under treatment condition $w$, we denote $T_i(w) = \infty$. Intercurrent events are considered as absent if no post-treatment intercurrent events occur until $t^*$, in which case we denote $R_i(w) = \infty$ or an arbitrary number larger than $t^*$, since only the events before $t^*$ will be counted in the analysis.

We will introduce estimands without the notation of censoring, as censoring refers to the problem of missing data and should not affect the definition of estimands. On the one hand, censoring does not change the potential risks of intercurrent events and primary outcome events under the commonly adopted random censoring assumption, saying that censoring is a nuisance. On the other hand, the potential times to intercurrent events and primary outcome events are still defined even if there is censoring. It simply means that we do not observe the outcomes if there is censoring. Censoring will be considered in Section \ref{sec3} when discussing identification and estimation. We adopt the potential cumulative incidences under both treatment assignments as the target estimands. Potential cumulative incidences describe the probability of time-to-event outcomes occurring at each time point. We define the treatment effect as the contrast of two potential cumulative incidences. Cumulative incidences are model-free and collapsible, enjoying causal interpretations. Other forms of treatment effects, such as restricted mean time lost, can also be derived by contrasting the integrated cumulative incidence functions \citep{zhao2016restricted, conner2021estimation}.

We will hereby suppress the individual $i$ subscript because the random vector for each individual is assumed to be drawn independently from a distribution common to all individuals. In the following, we use $\mathbb{P}(\cdot)$ to denote the probability taken with randomized sampling and treatment assignment, $\mathbb{E}(\cdot)$ to denote the expectation, and $\mathbb{I}(\cdot)$ to denote the indicator function.
We let $\mu_1^{k}(t)$ and $\mu_0^{k}(t)$, $k \in \{\text{tp, cv, hp, wo, ps}\}$ , $0 \leq t \leq t^*$, denote the two cumulative incidences of potential outcome events under the treatment policy strategy, composite variable strategy, hypothetical strategy, while on treatment strategy and principal stratum strategy, respectively. The mathematical forms of $\mu_1^{k}(t)$ and $\mu_0^{k}(t)$ will be illustrated below.

\paragraph{Treatment policy strategy}

The treatment policy strategy addresses the problem of intercurrent events by expanding the initial treatment conditions to a treatment policy. This strategy is applicable only if intercurrent events do not hinder primary outcome events; in this study, it is confined to a semi-competing risks context. The treatments under comparison are now two treatment policies: $(w, R(w))$, where $w = 1$ or $0$. One policy $(1,R(1))$ involves administering the test drug, along with any naturally occurring intercurrents, whereas the other policy $(0,R(0))$ involves administering a placebo, along with any naturally occurring intercurrents. Thus, the potential outcomes are $T(1,R(1))$ and $T(0,R(0))$. Instead of comparing the test drug and placebo themselves, the contrast of interest is made between the two treatment policies. The difference in cumulative incidences under the two treatment policies is then
\begin{equation}\label{esmd:diff:tp}
\begin{aligned}
\tau^{\text{tp}}(t) &:= \mu_1^{\text{tp}}(t) - \mu_0^{\text{tp}}(t) \\
&:= \mathbb{P}(T(1, R(1)) < t) - \mathbb{P}(T(0, R(0)) < t),
\end{aligned}
\end{equation}
representing the difference in probabilities of experiencing primary outcome events during $[0, t)$ under active treatment and placebo.

The average treatment effect $\tau^{\text{tp}}(t)$ has a meaningful causal interpretation only when $T(1, R(1))$ and $T(0, R(0))$ are well defined. Because the treatment policy includes the occurrence of the intercurrent event as natural, the entire treatment policy is determined by manipulating the initial treatment condition $w$ only. Therefore, we can simplify the notations $T(w, R(w)) = T(w)$ in defining estimands, $w =1, 0$. As such, $\tau^{\text{tp}}(t) = \mathbb{P}(T(1)) < t) - \mathbb{P}(T(0) < t)$ as the intention-to-treat analysis. 

\paragraph{Composite variable strategy}

The composite variable strategy addresses the problem of intercurrent events by expanding the outcome variables. It aggregates the intercurrent event and primary outcome event into a single composite outcome variable. The idea is not new in the context of progression-free survival \citep{broglio2009detecting, fleming2009issues}, where the composite outcome variable is defined as the occurrence of either a non-terminal event (e.g., cancer progression) or a terminal event (e.g., death). One widely used composite outcome variable has the form $R(w) \wedge T(w) = \min\{T(w), R(w)\}$ for $w \in \{1,0\}$. When this simple form is adopted, the difference in counterfactual cumulative incidences is
\begin{equation}\label{esmd:diff:cv}
\begin{aligned}
\tau^{\text{cv}}(t) &:= \mu_1^{\text{cv}}(t) - \mu_0^{\text{cv}}(t) \\
&:= \mathbb{P}( R(1) \wedge T(1) < t ) - \mathbb{P}( R(0) \wedge T(0) < t ),
\end{aligned}
\end{equation}
representing the difference in probabilities of experiencing either intercurrent events or primary outcome events during $[0, t)$ under active treatment and placebo.
Other composite forms of $T(w)$ and $R(w)$ are also available, e.g., the composite variable as quality-adjusted lifetime which imposes different weights for the pre-intercurrent-event period $T(w) \wedge R(w)$ and post-intercurrent-event period $T(w) - T(w) \wedge R(w)$ \citep{goldhirsch1989costs, cole1996quality, revicki2006analyzing}.

\paragraph{While on treatment strategy}

The while on treatment strategy considers the measure of outcome variables taken only up to the occurrence of intercurrent events. The failures of primary outcome events should not be counted in the cumulative incidences $\mu_w^{\text{wo}}(t)$, $w = 1, 0$, if intercurrent events occurred. The difference in counterfactual cumulative incidences under this strategy is
\begin{equation}\label{esmd:diff:wo2}
\begin{aligned}
\tau^{\text{wo}}(t) &:= \mu_1^{\text{wo}}(t) - \mu_0^{\text{wo}}(t) \\
&:= \mathbb{P}(T(1) < t, R(1) \geq t) - \mathbb{P}(T(0) < t, R(0) \geq t),
\end{aligned}
\end{equation}
representing the difference in probabilities of experiencing primary outcome events without intercurrent events during $[0, t)$ under active treatment and placebo. The $\mu_w^{\text{wo}}(t)$ is also known as the cause-specific cumulative incidence or subdistribution function \citep{geskus2011cause, lambert2017flexible}.

The while on treatment strategy is closely related to the competing risks model, which is widely used in medical studies \citep{fine1999proportional, lau2009competing, andersen2012competing}. However, for causal interpretations, it is worth emphasizing that the hazard of $R(1)$ may differ from that of $R(0)$, leading to vast difference in the underlying features of individuals who have not experienced the primary outcome event between treatment conditions until any time $t \in \{0,t^*]$. When the scientific question of interest is the impact of treatment on the primary outcome event, the estimand $\tau^{\text{wo}}(t)$ is hard to interpret if systematic difference in the risks of intercurrent events between two treatment conditions under comparison is anticipated \citep{latouche2013competing, han2023defining}.

\paragraph{Hypothetical strategy}\label{sec:hp:estimand}

The hypothetical strategy envisions a hypothetical clinical trial condition where the occurrence of intercurrent events is restricted in certain ways. By doing so, the distribution of potential outcomes under the hypothetical scenario can capture the impact of intercurrent events explicitly through a pre-specified criterion. We use $T'(w)$, $w = 1, 0$ to denote the time to the primary outcome event in the hypothetical scenario. The time-dependent treatment effect specific to this hypothetical scenario is written as
\begin{equation}\label{esmd:diff:hp}
\begin{aligned}
\tau^{\text{hp}}(t) &:= \mu_1^{\text{hp}}(t) - \mu_0^{\text{hp}}(t) \\
&:= \mathbb{P}(T'(1) < t) - \mathbb{P}(T'(0) < t),
\end{aligned}
\end{equation}
representing the difference in probabilities of experiencing primary outcome events during $[0, t)$ in the pre-specified hypothetical scenario under active treatment and placebo.

The key question is how to envision $T'(1)$ and $T'(0)$. For the competing risks data structure, there can be many hypothetical scenarios envisioned by manipulating the hazard specific to intercurrent event
\[
d\Lambda_2(t; w) = \mathbb{P}(t \leq R(w) < t+dt \mid T(w) \geq t, R(w) \geq t),
\]
while assuming the hazard specific to the primary outcome event
\[
d\Lambda_1(t; w) = \mathbb{P}(t \leq T(w) < t+dt \mid T(w) \geq t, R(w) \geq t)
\]
remains unchanged, where $w = 1, 0$ and $0 \leq t \leq t^*$. Let $d\Lambda_2'(t;w)$ and $d\Lambda_1'(t;w)$ be the hazards specific to the intercurrent event and primary outcome event in the hypothetical scenario respectively, with $d\Lambda_1'(t;w) = d\Lambda_1(t;w)$, $w = 1, 0$. For example, we can envision a scenario where the intercurrent events that occurred when individuals were assigned to test drugs were only permitted if these intercurrent events would have also occurred if these individuals had been assigned to the placebo. In this hypothetical scenario, when assigned to placebo, individuals would be equally likely to experience intercurrent events as they are assigned to placebo in the real-world trial in terms of the hazards; when assigned to test drug, the hazard of intercurrent events would be identical to that if assigned to placebo in the real-world trial. That is, $d\Lambda_2'(t;0) = d\Lambda_2'(t;1) = d\Lambda_2(t;0)$. We call this the hypothetical scenario I (corresponding to the nature effect), and denote the corresponding estimands as $\mu_j^{\text{hp,I}}(t)$ and $\tau^{\text{hp,I}}(t)$.
Alternatively, we can envision another hypothetical scenario where the intercurrent events are absent in the hypothetical scenario for all individuals, so $d\Lambda_2'(t;0) = d\Lambda_2'(t;1) = 0$ \citep{robins1992identifiability, young2020epidemiologic}. We call this the hypothetical scenario II (corresponding to the controlled effect with the intercurrent events removed), and denote the corresponding estimands as $\mu_j^{\text{hp,II}}(t)$ and $\tau^{\text{hp,II}}(t)$. This hypothetical scenario II leads to an estimand called the marginal cumulative incidence.

The hypothetical strategy is closely related to mediation analysis, which is intended to evaluate the direct and indirect effects for semi-competing risks data \citep{lange2011direct, martinussen2011estimation, tchetgen2011causal}. There are two hazards associated with the primary outcome event: one for the direct event following treatment and the other for the indirect event following intercurrent events. The invariance of the hazards specific to primary outcome events when manipulating the hazard specific to the intercurrent event is referred to as sequential ignorability in mediation analysis and is called dismissible components condition in the separable effects framework \citep{martinussen2023estimation, breum2024estimation, janvin2024causal, deng2024direct}. Under sequential ignorability or dismissible components condition, the hypothetical strategy estimand evaluates the direct effect of the treatment on the primary outcome event. The inteventional effect (or randomized, stochastic, organic) is another hypothetical strategy especially useful when there are time-varying confounders, assuming a random draw for the intercurrent event from some distribution in the hypothetical scenario \citep{vanderweele2017mediation, lin2017mediation}. Due to the complexity of estimation, we only focus the competing risks data structure here.



\paragraph{Principal stratum strategy}

The principal stratum strategy aims to stratify the population into subpopulations based on the joint potential occurrences of intercurrent events under the two treatment assignments $(R(1), R(0))$ \citep{frangakis2002principal, bornkamp2021principal, gao2023defining}. Suppose we are interested in a principal stratum comprised of individuals who would never experience intercurrent events, regardless of which treatment they receive. This principal stratum can be indicated by $\{R(1)=R(0)=\infty\}$. The treatment effect is now defined within this subpopulation,
\begin{equation}\label{esmd:diff:ps}
\begin{aligned}
\tau^{\text{ps}}(t) &:= \mu_1^{\text{ps}}(t) - \mu_0^{\text{ps}}(t) \\
&:= \mathbb{P}(T(1) < t \mid R(1)=R(0)=\infty) 
- \mathbb{P}(T(0) < t \mid R(1)=R(0)=\infty),
\end{aligned}
\end{equation}
representing the difference in probabilities of experiencing primary outcome events during $[0, t)$ under active treatment and placebo in the subpopulation that will not experience intercurrent events regardless of treatment during $[0, t)$.
The principal stratum strategy is especially useful when we have no accurate measurement of intercurrent events. In some clinical trials, we may only know whether an individual has experienced intercurrent events by the end of the study, but not know the exact time at which these events occurred. The principal stratum $\{R(1)=R(0)=\infty\}$ and the principal stratum estimand are still meaningful.

More generally, the principal stratum of interest can be defined as any indicator function of $(R(1),R(0),T(1),T(0))$ and time $t$ \citep{mattei2024assessing}. For example, the principal stratum $\{R(1)=\infty\}$ represents the subpopulation that will not experience intercurrent events under the active treatment. By allowing the principal stratum $\{R(1)>t, R(0)>t\}$ to depend on $t$, this principal stratum represents the subpopulation that will not experience intercurrent events, regardless of treatment, by time $t$.
However, the target population is impossible to identify, not only because values of $R(1)$ and $R(0)$ are never observed simultaneously, but also due to the problem of censoring so that whether $R(w)$ equals $\infty$ is unknown even under the assigned treatment condition $w$. The interpretation of the target population should be paid special attention because the composition of the target population relies on the choice of $t^*$ \citep{comment2019survivor}. The target population would decrease with increasing $t^*$, except if there is an upper limit, which is smaller than $t^*$ almost surely, for the occurrence time of intercurrent events. Typically, untestable assumptions are involved to identify the principal stratum estimand. Sensitivity analysis is recommended in conjunction with the principal stratum strategy to assess the robustness of conclusions to untestable assumptions \citep{gilbert2003sensitivity, stuart2015assessing, wang2023sensitivity}.

\section{Identification and estimation} \label{sec3}

\subsection{Data structure and notations with censoring}

Before describing the observed data structure, we update our notations to account for censoring. Each individual is associated with two potential right-censoring times, $C(1)$ and $C(0)$. Because of censoring, we are only able to partially observe potential primary outcome events through $\Delta^T(w) = \mathbb{I}\{T(w) \leq C(w)\}$ and $\tilde{T}(w) = T(w) \wedge C(w)$, and potential intercurrent events through $\Delta^R(w) = \mathbb{I}\{R(w) \leq T(w) \wedge C(w)\}$ and $\tilde{R}(w) = R(w) \wedge T(w) \wedge C(w)$. Let $\tilde{T}$ be the observed time to the primary outcome event (or censoring) with event indicator $\Delta^T$, and $\tilde{R}$ be the observed time to the intercurrent event (or censoring) with event indicator $\Delta^R$. We assume the stable unit treatment value assumption (SUTVA) and causal consistency as follows \citep{rubin1980randomization}.

\begin{assumption}[SUTVA] \label{asm:sut}
All individuals are independent, and there is only one version of potential outcomes $(T(w), R(w), C(w))$ associated with each treatment condition $w \in \{0,1\}$.
\end{assumption}
\begin{assumption}[Consistency] \label{asm:cons}
$\tilde{T} = \tilde{T}(W)$, $\Delta^T = \Delta^T(W)$, $\tilde{R} = \tilde{R}(W)$, and $\Delta^R = \Delta^R(W)$.
\end{assumption}

The observed data structure can be simplified according to how intercurrent events compete with primary outcome events.  If the intercurrent events do not prevent the occurrence of the primary outcome events (semi-competing risks data), our observed data has a structure of $(W, \tilde{T}, \tilde{R}, \Delta^T, \Delta^R)$. On the other hand, if intercurrent events can prevent the occurrence of primary outcome events (competing risks data), our observed data structure takes the form $(W, \tilde{T}\wedge\tilde{R}, \Delta^T, \Delta^R)$. For competing risks data, the event indicators are usually recoded as a single value $J$, where $J = 1$ if the primary outcome event is observed, $J = 2$ if the intercurrent event is observed, and $J = 0$ if the censoring event is observed.

To identify the causal estimands described in Section \ref{sec2}, we list the following four assumptions.

\begin{assumption}[Randomization] \label{asm:ran}
$W \perp (T(1), T(0), R(1), R(0))$.
\end{assumption}
\begin{assumption}[Random censoring] \label{asm:non}
$C(w) \perp (T(w), R(w)) \mid W = w$, for $w = 1, 0$.
\end{assumption}
\begin{assumption}[Positivity] \label{asm:pos}
$ \mathbb{P}(T(w) > t^*, R(w) > t^*, C^*(w) > t^*, W = w) > 0$, for $w = 1, 0$.
\end{assumption}
\begin{assumption}[Principal ignorability] \label{asm:pri}
$T(w) \perp R(1-w) \mid R(w)$, for $w = 1, 0$.
\end{assumption}

Assumptions \ref{asm:sut}--\ref{asm:pos} should be made under all these five strategies for identification, and Assumption \ref{asm:pri} is added under the principal stratum strategy. Although the principal stratum estimand can be identified under other assumptions instead of Assumption \ref{asm:pri} (for example, conditional principal ignorability on baseline covariates as well as monotonicity \citep{feller2017principal, ding2017principal}), we adopt Assumption \ref{asm:pri} here because it has the most straightforward form in principal sratification.
Below, we will focus on the hazard functions of events to identify the target causal estimands, as the random censoring assumption ensures that missingness is at random given at-risk sets at each time point.

\subsection{Treatment policy strategy}

\subsubsection{Identification}

Under Assumptions \ref{asm:sut}--\ref{asm:pos}, the hazard function of the potential primary outcome event can be identified through
\begin{align*}
d\Lambda(t;w) &:= \mathbb{P}(t \leq T(w) < t+dt \mid T(w) \ge t) \\
&= \mathbb{P}(t \leq \tilde{T}(w) < t+dt, \Delta^T(w) = 1 \mid \tilde{T}(w) \ge t), \\
&= \mathbb{P}(t \leq \tilde{T} < t+dt, \Delta^T = 1 \mid \tilde{T} \ge t, W = w),
\end{align*}
$w = 1, 0$. Because
\begin{align*}
\mu_w^{\text{tp}}(t) &= \mathbb{P}(T(w) < t) = 1- \exp\{-\Lambda(t;w)\},
\end{align*} we can identify $\tau^{\text{tp}}(t)$ as
\begin{align*}
\tau^{\text{tp}}(t)
&= \exp\{-\Lambda(t;0)\} - \exp\{-\Lambda(t;1)\}.
\end{align*}

\subsubsection{Estimation}

To estimate this estimand, we need the data on the primary outcome events, $(\tilde{T}, \Delta^T, W)$. We write the event process and at-risk process of the primary outcome event as follows,
\begin{align*}
N(t;w) &= \sum_{i=1}^{n} \mathbb{I}\{\tilde{T}_i \leq t, \Delta^T_i = 1, W_i=w\}, \\
Y(t;w) &= \sum_{i=1}^{n} \mathbb{I}\{\tilde{T}_i \geq t, W_i=w\}, \quad w = 1, 0.
\end{align*}
The cumulative hazard function $\Lambda(t;w)$, $w = 1, 0$ is then consistently and unbiasedly estimated by the Nelson--Aalen estimator
\[
\widehat\Lambda(t;w) = \int_0^t \frac{dN(s;w)}{Y(s;w)}.
\]

We shall estimate $\mu_w^{\text{tp}}(t)$ and $\tau^{\text{tp}}(t)$ by the plug-in estimator using the estimated hazard $\widehat\Lambda(t;w)$, denoted by $\widehat\mu_w^{\text{tp}}(t)$ and $\widehat\tau^{\text{tp}}(t)$ respectively. The pointwise asymptotic variances of $\widehat\mu_w^{\text{tp}}(t)$ and $\widehat\tau^{\text{tp}}(t)$ are
\begin{align*}
\avar\{\widehat\mu_w^{\text{tp}}(t)\} &= \exp\{-2\Lambda(t;w)\} \int_0^t \frac{d\Lambda(s;w)}{\mathbb{P}(\tilde{T} \ge s, W=w)}, \\
\avar\{\widehat\tau^{\text{tp}}(t)\} &= \avar\{\widehat\mu_1^{\text{tp}}(t)\} + \avar\{\widehat\mu_0^{\text{tp}}(t)\},
\end{align*}
which can be consistently estimated by plug-in estimators. Thus, the confidence intervals of $\mu_w^{\text{tp}}(t)$ and $\tau^{\text{tp}}(t)$ can be constructed based on the normal approximation.

\subsection{Composite variable strategy}

\subsubsection{Identification}

Under Assumptions \ref{asm:sut}--\ref{asm:pos}, the hazard function of the composite outcome variable, the potential minimum event time $T(w) \wedge R(w)$, can be identified through
\begin{align*}
d\Lambda_{12}(t;w) &:= \mathbb{P}(t \leq T(w) \wedge R(w) < t+dt \mid T(w) \wedge R(w) \ge t) \\
&= \mathbb{P}(t \leq \tilde{T} \wedge \tilde{R} < t+dt, \Delta^T \vee \Delta^R = 1 \mid \tilde{T} \wedge \tilde{R} \ge t, W = w),
\end{align*}
$w=1,0$, where $\Delta^T \vee \Delta^R = \max\{\Delta^T, \Delta^R\}$. Under the treatment condition $w$, the probability that an individual experiences a composite outcome event before $t$ is $\mu_w^{\text{cv}}(t) = 1-\exp\{-\Lambda_{12}(t;w)\}$. The estimand $\tau^{\text{cv}}(t)$ is identified as
\[
\tau^{\text{cv}}(t) = \exp\{-\Lambda_{12}(t;0)\} - \exp\{-\Lambda_{12}(t;1)\}.
\]

\subsubsection{Estimation}

To estimate this estimand, we collect data on the earlier one of the primary outcome event and intercurrent event. The data have the structure of $(\tilde{T} \wedge \tilde{R}, \Delta^T \vee \Delta^R, W)$. Similarly, we will focus on estimating hazards and then use the plug-in estimators to estimate $\mu_w^{\text{cv}}(t)$ and $\tau^{\text{cv}}(t)$.

We write the event process and at-risk process of the composite outcome event as follows,
\begin{align*}
N_{12}(t;w) &= \sum_{i=1}^{n} \mathbb{I}\{\tilde{T}_i \wedge \tilde{R}_i \leq t, \Delta^T_i \vee \Delta^R_i = 1, W_i=w\},  \\
Y_{12}(t;w) &= \sum_{i=1}^{n} \mathbb{I}\{\tilde{T}_i \wedge \tilde{R}_i \geq t, W_i=w\}.
\end{align*}
The cumulative hazard function $\Lambda_{12}(t;w)$, $w = 1, 0$ is consistently and unbiasedly estimated by
\[
\widehat\Lambda_{12}(t;w) = \int_0^t \frac{dN_{12}(s;w)}{Y_{12}(s;w)}.
\]
The pointwise asymptotic variances of the plug-in estimators of $\mu_w^{\text{cv}}(t)$ and $\tau^{\text{cv}}(t)$ are given by
\begin{align*}
\avar\{\widehat\mu_w^{\text{cv}}(t)\} &= \exp\{-2\Lambda_{12}(t;w)\} \int_0^t \frac{d\Lambda_{12}(s;w)}{\mathbb{P}(\tilde{T}\wedge\tilde{R}\ge s, W=w)}, \\
\avar\{\widehat\tau^{\text{cv}}(t)\} &= \avar\{\widehat\mu_1^{\text{cv}}(t)\} + \avar\{\widehat\mu_0^{\text{cv}}(t)\}.
\end{align*}

\subsection{While on treatment strategy}\label{wo:estimation}

\subsubsection{Identification}

We denote the hazard specific to the primary outcome event and the hazard specific to the intercurrent event by
\begin{align*}
d\Lambda_1(t;w) &= \mathbb{P}(t \leq T(w) < t+dt \mid T(w) \geq t, R(w) \geq t), \\
d\Lambda_2(t;w) &= \mathbb{P}(t \leq R(w) < t+dt \mid T(w) \geq t, R(w) \geq t),
\end{align*}
respectively, where $w = 1, 0$.
Under Assumptions \ref{asm:sut}--\ref{asm:pos}, they are identified as
\begin{align*}
d\Lambda_1(t;w) 
&= \mathbb{P}(t \leq \tilde{T} < t+dt, \Delta^T = 1 \mid \tilde{T} \geq t, \tilde{R} \geq t, W = w), \\
d\Lambda_2(t;w) 
&= \mathbb{P}(t \leq \tilde{R} < t+dt, \Delta^R = 1 \mid \tilde{T} \geq t, \tilde{R} \geq t, W = w).
\end{align*}
The summation of these two hazards, $d\Lambda_1(t;w) + d\Lambda_2(t;w)$, is the hazard of developing either primary outcome event or intercurrent event, which is the same as the hazard under the composite variable outcome, that is,
\[
d\Lambda_1(t;w) + d\Lambda_2(t;w) = d\Lambda_{12}(t;w).
\]
Because
\begin{align*}
\mu_w^{\text{wo}}(t) &= \mathbb{P}(T(w) < t, R(w) \geq t)
= \int_0^t  \exp\{-\Lambda_{12}(s;w)\} d\Lambda_1(s;w),
\end{align*}
$w = 1, 0$, the estimand $\tau^{\text{wo}}(t)$ is identified as
\begin{align*}
\tau^{\text{wo}}(t) &= \int_0^t \exp\{-\Lambda_{12}(s;1)\}d\Lambda_1(s;1) - \int_0^t \exp\{-\Lambda_{12}(s;0)\}d\Lambda_1(s;0).
\end{align*}

\subsubsection{Estimation}

Similarly, we write the event process of the primary outcome event, event process of the intercurrent event and at-risk process as follows,
\begin{align*}
N_1(t;w) &= \sum_{i=1}^{n} \mathbb{I}\{\tilde{T}_i \leq t, \Delta^T_i = 1, W_i=w\}, \\ N_2(t;w) &= \sum_{i=1}^{n} \mathbb{I}\{\tilde{R}_i \leq t, \Delta^R_i = 1, W_i=w\}, \\
Y_{12}(t;w) &= \sum_{i=1}^{n} \mathbb{I}\{\tilde{T}_i \geq t, \tilde{R}_i \geq t, W_i=w\}.
\end{align*}
The cumulative hazard function $\Lambda_j(t;w)$, $j = 1, 2$, is consistently and unbiasedly estimated by
\[
\widehat\Lambda_j(t;w) = \int_0^t \frac{dN_j(s;w)}{Y_{12}(s;w)}.
\]
The pointwise asymptotic variances of the plug-in estimators of $\mu_w^{\text{wo}}(t)$ and $\tau^{\text{wo}}(t)$ are given by
\begin{align*}
\avar\{\widehat\mu_w^{\text{wo}}(t)\} &=
\int_0^t \bigg[ \{e^{-\Lambda_{12}(s;w)}-\mu_w^{\text{wo}}(t)+\mu_w^{\text{wo}}(s)\}^2 \frac{d\Lambda_1(s;w)}{\mathbb{P}(\tilde{T}\wedge\tilde{R} \ge s, W=w)} \\
&\qquad\quad + \{\mu_w^{\text{wo}}(t)-\mu_w^{\text{wo}}(s)\}^2 \frac{d\Lambda_2(s;w)}{\mathbb{P}(\tilde{T}\wedge\tilde{R} \ge s, W=w)} \bigg], \\
\avar\{\widehat\tau^{\text{wo}}(t)\} &= \avar\{\widehat\mu_1^{\text{wo}}(t)\} + \avar\{\widehat\mu_0^{\text{wo}}(t)\}.
\end{align*}
To estimate these quantities, we need to collect the data $(\tilde{T} \wedge \tilde{R}, \Delta^T, \Delta^R, W)$.

\subsection{Hypothetical strategy}\label{estimation:hp}

We consider two hypothetical scenarios as described in Section \ref{sec:hp:estimand}, the hypothetical scenario I where the intercurrent events when individuals were assigned to the test drug were only permitted if these intercurrent events would have occurred if these individuals were assigned to placebo, and hypothetical scenario II, where the intercurrent events are absent in the hypothetical scenario for all individuals. Due to the complexity of estimation and inference under semi-competing risks, we only consider the competing risks data structure in this section. Nonparametric estimation and inference for the semi-competing risks data structure are discussed in literature \citep{huang2021causal, deng2024direct}.

\subsubsection{Identification}

In the hypothetical scenario, under treatment condition $w$, the primary outcome event is generated according to the hazard $d\Lambda_1'(t;w)$ and the intercurrent event is generated according to the hazard $d\Lambda_2'(t;w)$. Due to competing risks, the cumulative incidence of the potential primary outcome event at time $t$ in the hypothetical scenario is
\begin{align*}
\mu_j^{\text{hp}}(t)
&= \int_0^t \exp\{-\Lambda_1'(s;w) - \Lambda_2'(s;w)\}d\Lambda_1'(s;w).
\end{align*}
Under Assumptions \ref{asm:sut}--\ref{asm:pos}, the hazards $d\Lambda_1(t;w)$ and $d\Lambda_2(t;w)$ are identifiable. The estimand of hypothetical scenario I can be identified as follows,
\begin{align*}
\tau^{\text{hp,I}}(t)
&= \int_0^t \exp\{-\Lambda_1(s;1) - \Lambda_2(s;0)\}d\Lambda_1(s;1) - \int_0^t \exp\{-\Lambda_{1}(s;0) - \Lambda_{2}(s;0)\}d\Lambda_1(s;0).
\end{align*}
Note that $\mu_0^{\text{hp,I}}(t)$ is identical to $\mu_0^{\text{wo}}(t)$ in the while on treatment strategy in Section \ref{wo:estimation}. Within the mediation analysis framework, $\tau^{\text{hp,I}}(t)$ represents the natural direct effect on the cumulative incidence of the primary outcome event under sequential ignorability, controlling for the hazard of the intercurrent event at the level of $w=0$ \citep{deng2024direct}. Formally, sequential ignorability assumes that the hazard of the primary outcome event (or intercurrent event) under treatment condition $w$ in the hypothetical world does not depend on the cross-world ($1-w$) hazard of the intermediate event (or primary outcome event). Sequential ignorability has a similar interpretation to principal ignorability: There is no common cause for the potential primary outcome event and intercurrent event.

Similarly, the estimand under the hypothetical scenario II can be identified as follows,
\begin{align*}
\tau^{\text{hp,II}}(t)
&= \int_0^t \exp\{-\Lambda_1(s;1)\}d\Lambda_1(s;1) - \int_0^t \exp\{-\Lambda_1(s;0)\}d\Lambda_1(s;0).
\end{align*}
Of note, $\tau^{\text{hp,II}}(t)$ evaluates the contrast of the marginal distributions of $T(1)$ and $T(0)$ with intercurrent events viewed as independent censoring \citep{emura2020comparison}. Within the mediation analysis framework, $\tau^{\text{hp,II}}(t)$ represents the controlled direct effect on the primary outcome event where the hazard specific to intercurrent events is controlled at zero under sequential ignorability.

\subsubsection{Estimation}

As in the previous section, we estimate and infer these two hypothetical strategy estimands by plugging the estimates $\widehat\Lambda_j(t;w)$, $ j = 1, 2, 12$, which requires the data of $(\tilde{T} \wedge \tilde{R}, \Delta^T, \Delta^R, W)$. The pointwise asymptotic variances of the plug-in estimators of cumulative incidences and treatment effects are
\begin{align*}
\avar\{\widehat\mu_w^{\text{hp,I}}(t)\}
&= \int_0^t \{e^{-\Lambda_1(s;w)-\Lambda_2(s;0)}-\mu_w^{\text{hp,I}}(t)+\mu_w^{\text{hp,I}}(s)\}^2 \frac{d\Lambda_1(s;w)}{\mathbb{P}(\tilde{T}\wedge\tilde{R} \ge s, W=1)} \\
&\qquad + \int_0^t \{\mu_w^{\text{hp,I}}(t)-\mu_w^{\text{hp,I}}(s)\}^2 \frac{d\Lambda_2(s;0)}{\mathbb{P}(\tilde{T}\wedge\tilde{R} \ge s, W=0)}, \\
\avar\{\widehat\tau^{\text{hp,I}}(t)\}
&= \int_0^t \{e^{-\Lambda_1(s;1)-\Lambda_2(s;0)}-\mu_1^{\text{hp,I}}(t)+\mu_1^{\text{hp,I}}(s)\}^2 \frac{d\Lambda_1(s;1)}{\mathbb{P}(\tilde{T}\wedge\tilde{R} \ge s, W=1)} \\
&\qquad + \int_0^t \{e^{-\Lambda_1(s;0)-\Lambda_2(s;0)}-\mu_0^{\text{hp,I}}(t)+\mu_0^{\text{hp,I}}(s)\}^2 \frac{d\Lambda_1(s;0)}{\mathbb{P}(\tilde{T}\wedge\tilde{R} \ge s, W=0)} \\
&\qquad + \int_0^t \{\mu_1^{\text{hp,I}}(t)-\mu_0^{\text{hp,I}}(t)-\mu_1^{\text{hp,I}}(s)+\mu_0^{\text{hp,I}}(s)\}^2 \frac{d\Lambda_2(s;0)}{\mathbb{P}(\tilde{T}\wedge\tilde{R} \ge s, W=0)}, \\
\avar\{\widehat\mu_w^{\text{hp,II}}(t)\}
&= \exp\{-2\Lambda_1(t;w)\} \int_0^t \frac{d\Lambda_1(s;w)}{\mathbb{P}(\tilde{T}\wedge\tilde{R} \ge s, W=w)}, \\
\avar\{\widehat\tau^{\text{hp,II}}(t)\} &= \avar\{\widehat\mu_1^{\text{hp,II}}(t)\} + \avar\{\widehat\mu_0^{\text{hp,II}}(t)\}.
\end{align*}
Here the expression for $\avar\{\widehat\tau^{\text{hp,I}}(t)\}$ is not the simple summation of $\avar\{\widehat\mu_1^{\text{hp,I}}(t)\}$ and $\avar\{\widehat\mu_0^{\text{hp,I}}(t)\}$ because $\widehat\mu_1^{\text{hp,I}}(t)$ and $\widehat\mu_0^{\text{hp,I}}(t)$ depend on a common factor $d\widehat\Lambda_2(t;0)$.

\subsection{Principal stratum strategy}\label{estimation:ps}

\subsubsection{Identification}

The target population of the principal stratum strategy $\{R(1) = R(0) = \infty\}$ involves the joint distribution of $R(1)$ and $R(0)$, which is not identifiable from observed data. We rely on Assumption \ref{asm:pri} to identify the principal stratum estimand $\tau^{\text{ps}}(t)$. Assumption \ref{asm:pri} states that the potential time to the primary outcome event $T(w)$ can be correlated with $R(w)$ but should not have cross-world reliance on $R(1-w)$. Therefore, the distribution of $T(w)$ is identical in the group $\{R(w) = \infty\}$ and in the principal stratum $\{R(1) = R(0) = \infty\}$. This assumption holds if there is no common cause for the potential intercurrent event and primary outcome event, similar to sequential ignorability. Recall that we have assumed that $R(w) = \infty$ if the individual would not experience intercurrent events before the end of study, so $R(w) = \infty$ is equivalent to $R(w) > T(w)$, and also $R(w) > t^*$. Thus,
\begin{align*}
\tau^{\text{ps}}(t) &= \mathbb{P}(T(1) < t \mid R(1) = R(0) = \infty) - \mathbb{P}(T(0) < t \mid R(1) = R(0) = \infty) \\
&= \mathbb{P}(T(1) < t \mid R(1) = \infty) - \mathbb{P}(T(0) < t \mid R(0) = \infty) \\
&= \frac{\mathbb{P}(T(1) < t, R(1) = \infty)}{\mathbb{P}(R(1) = \infty)} - \frac{\mathbb{P}(T(0) < t, R(0) = \infty)}{\mathbb{P}(R(0) = \infty)} \\
&= \frac{\mathbb{P}(T(1) < t, R(1) \geq t)}{\mathbb{P}(R(1) > t^*)} - \frac{\mathbb{P}(T(0) < t, R(0) \geq t)}{\mathbb{P}(R(0) > t^*)}.
\end{align*}
The second equation is by principal ignorability, and the fourth equation is by the definition of intercurrent events. The numerators can be identified in the same way as the estimand in the while on treatment strategy under Assumptions \ref{asm:sut}--\ref{asm:pos}. Since
\[
\mathbb{P}(R(w) < t^*) = \mathbb{P}(R(w) < t^*, T(w) > R(w)) = \int_0^{t^*} \exp\{-\Lambda_{12}(s;w)\} d\Lambda_2(s;w),
\]
we have
\[
\mu_w^{\text{ps}}(t) = \frac{\int_0^t \exp\{-\Lambda_{12}(s;w)\} d\Lambda_1(s;w)}{1 - \int_0^{t^*} \exp\{-\Lambda_{12}(s;w)\} d\Lambda_2(s;w)},
\]
$w = 1, 0$.
Because $\mathbb{P}(R(w) > t^*) \leq 1$, the following inequality always holds:
\begin{align*}
\mu_w^{\text{ps}}(t) &= \frac{\mathbb{P}(T(1) < t, R(1) \geq t)}{\mathbb{P}(R(1) > t^*)} \geq \mathbb{P}(T(1) < t, R(1) \geq t) = \mu_w^{\text{wo}}(t).
\end{align*}
The cumulative incidences under the principal stratum strategy are always larger than those under the while on treatment strategy.

\subsubsection{Estimation}

Similarly, we estimate the estimand $\tau^{\text{ps}}(t)$ by plugging the estimates $\widehat\Lambda_j(t;w)$, $j = 1, 2, 12$ in the expression of $\tau^{\text{ps}}(t)$. Data of structure $(\tilde{T} \wedge \tilde{R}, \Delta^T, \Delta^R, W)$ is needed.

Utilizing the functional delta method, the pointwise asymptotic variances of the plug-in estimators of $\mu_w^{\text{ps}}(t)$ and $\tau^{\text{ps}}(t)$ are given by (see Supplementary Material A)
\begin{align*}
\avar\{\widehat\mu_w^{\text{ps}}(t)\}
&= \frac{\int_0^{t^*} \{A_1(s,t,w)-\mu_w^{\text{ps}}(t)A_2(s,t^*,w)\}^2 \mathbb{P}(\tilde{T}\wedge\tilde{R} \ge s, W=w)^{-1}d\Lambda_1(s;w)}{\{1 - \int_0^{t^*} e^{-\Lambda_{12}(s;w)} d\Lambda_2(s;w)\}^2} \\
&\quad + \frac{\int_0^{t^*} \{B_1(s,t,w)-\mu_w^{\text{ps}}(t)B_2(s,t^*,w)\}^2 \mathbb{P}(\tilde{T}\wedge\tilde{R} \ge s, W=w)^{-1}d\Lambda_2(s;w)}{\{1 - \int_0^{t^*} e^{-\Lambda_{12}(s;w)} d\Lambda_2(s;w)\}^2}, \\
\avar\{\widehat\tau^{\text{ps}}(t)\} &= \avar\{\widehat\mu_1^{\text{ps}}(t)\} + \avar\{\widehat\mu_0^{\text{ps}}(t)\},
\end{align*}
where
\begin{align*}
A_1(s,t,w) &= [\exp\{-\Lambda_{12}(s;w)\}+\mu_w^{\text{wo}}(s)-\mu_w^{\text{wo}}(t)] I\{s \le t\}, \\
A_2(s,t,w) &= \exp\{-\Lambda_{12}(s;w)\}-\exp\{-\Lambda_{12}(t;w)\}+\mu_w^{\text{wo}}(s)-\mu_w^{\text{wo}}(t), \\
B_1(s,t,w) &= \{\mu_w^{\text{wo}}(t)-\mu_w^{\text{wo}}(s)\} I\{s \le t\}, \\
B_2(s,t,w) &= \exp\{-\Lambda_{12}(t;w)\}+\mu_w^{\text{wo}}(t)-\mu_w^{\text{wo}}(s).
\end{align*}

The key idea to estimate $\mu_w^{\text{ps}}(t)$ is to estimate the proportion of the target principal stratum and the cumulative incidence function of the primary outcome event. We estimate the proportion of the target principal stratum by modeling the hazard of the intercurrent event. Using the occurrence indicators rather than occurrence times of intercurrent events, a mixture model can be employed, which consists of a submodel for the principal stratum proportion and a submodel for the cumulative incidence function of the primary outcome event. The EM algorithm is usually used to fit this mixture model \citep{gao2023defining, mattei2024assessing, liu2024principal}.

A typical way to relax the principal ignorability assumption is to assume conditional principal ignorability \citep{feller2017principal, ding2017principal}. At each level of baseline covariates $X$, the conditional principal stratum estimand is identifiable. However, it is challenging to integrate covariates out from the conditional principal stratum $\{R(1)=R(0)=\infty, X\}$ to the unconditional principal stratum $\{R(1)=R(0)=\infty\}$, where the covariates shift leads to a weight proportional to $\mathbb{P}(R(1)=R(0)=\infty \mid X)$. Under the monotonicity that each individual has better tolerance of intercurrent events under active treatment than under placebo $R(1) \geq R(0)$, the weight becomes $\mathbb{P}(R(0)=\infty \mid X)$ and hence the target estimand is identifiable. However, it is worth conducting sensitivity analyses because the monotonicity assumption may not hold \citep{lee2010causal, small2017instrumental}.

\section{Hypothesis testing} \label{sec4}

In this section, we conduct hypothesis testing on the treatment effects. Under the treatment policy strategy, the null and alternative hypotheses are
\begin{align*}
&H_0^{\text{tp}}: \tau^{\text{tp}}(t) = 0, ~\forall~t<t^*  \\
\text{ versus } ~& H_1^{\text{tp}}: \tau^{\text{tp}}(t) \neq 0, ~\exists~t<t^*,
\end{align*}
which is equivalent to
\begin{align*}
&H_0^{\text{tp}}: \Lambda(t;1) = \Lambda(t;0), ~\forall~t<t^* \\
\text{ versus } ~& H_1^{\text{tp}}: \Lambda(t;1) \neq \Lambda(t;0), ~\exists~t<t^*,
\end{align*}
testing whether the hazards of the primary outcome event with intercurrent events as natural are identical under active treatment and placebo.

The hypothesis test under the composite variable strategy tests is
\begin{align*}
&H_0^{\text{cv}}: \tau^{\text{cv}}(t) = 0, ~\forall~t<t^* \\
\text{ versus } ~& H_1^{\text{cv}}: \tau^{\text{cv}}(t) \neq 0, ~\exists~t<t^*,
\end{align*}
which is equivalent to
\begin{align*}
&H_0^{\text{cv}}: \Lambda_{12}(t;1) = \Lambda_{12}(t;0), ~\forall~t<t^* \\
\text{ versus } ~& H_1^{\text{cv}}: \Lambda_{12}(t;1) \neq \Lambda_{12}(t;0), ~\exists~t<t^*,
\end{align*}
testing whether the hazards of the composite of the primary outcome event and intercurrent event are identical under active treatment and placebo.

The hypothesis test under the hypothetical strategy is
\begin{align*}
&H_0^{\text{hp}}: \tau^{\text{hp}}(t) = 0, ~\forall~t<t^* \\
\text{ versus } ~& H_1^{\text{hp}}: \tau^{\text{hp}}(t) \neq 0, ~\exists~t<t^*.
\end{align*}
Because the hazards of the intercurrent event are controlled in hypothetical scenarios I and II, only the hazards of the primary outcome event are relevant. So the test is equivalent to
\begin{align*}
&H_0^{\text{hp}}: \Lambda_{1}(t;1) = \Lambda_1(t;0), ~\forall~t<t^* \\
\text{ versus } ~& H_1^{\text{hp}}: \Lambda_1(t;1) \neq \Lambda_1(t;0), ~\exists~t<t^*,
\end{align*}
testing whether the hazards of the primary outcome event are identical under the active treatment and placebo.

Analytical forms of hypothesis testing under the while on treatment strategy and principal stratum strategy are not available using simple log-rank tests because the hazards under these strategies involve more than one hazard function. Although test statistics of other forms are possible (for example, the Gray test for the while on treatment strategy), the interpretation of such test statistics may not be intuitive \citep{gray1988class}. As such, we shall focus on the above three hypothesis tests only, where log-rank tests can be applied. For any left-continuous predictable weight function $\omega(t)$ possibly depending on observed data, the log-rank test statistics are given by
\begin{align*}
U^{\text{tp}} &= \int_0^{t^*} \omega(s) \frac{Y(s;1)dN(s;0)-Y(s;0)dN(s;1)}{Y(s;1)+Y(s;0)}, \\
U^{\text{cv}} &= \int_0^{t^*} \omega(s) \frac{Y_{12}(s;1)dN_{12}(s;0)-Y_{12}(s;0)dN_{12}(s;1)}{Y_{12}(s;1)+Y_{12}(s;0)}, \\
U^{\text{hp}} &= \int_0^{t^*} \omega(s) \frac{Y_{12}(s;1)dN_1(s;0)-Y_{12}(s;0)dN_1(s;1)}{Y_{12}(s;1)+Y_{12}(s;0)}.
\end{align*}
In practice, we may set $ \omega(t) \equiv 1$. Under the null hypotheses $H_0^{\text{tp}}$, $H_0^{\text{cv}}$ and $H_0^{\text{hp}}$, the standardized test statistics follow asymptotic normal distribution,
\begin{align*}
U^{k}/\sqrt{S^{k}} \rightarrow_d N(0, 1), \quad k = \text{tp}, \text{cv}, \text{hp},
\end{align*}
with
\begin{align*}
S^{\text{tp}} &= \int_0^{t^*} \omega(s)^2 \frac{Y(s;1)Y(s;0)\{dN(s;1)+dN(s;0)\}}{\{Y(s;1)+Y(s;0)\}^2}, \\
S^{\text{cv}} &= \int_0^{t^*} \omega(s)^2 \frac{Y_{12}(s;1)Y_{12}(s;0)\{dN_{12}(s;1)+dN_{12}(s;0)\}}{\{Y_{12}(s;1)+Y_{12}(s;0)\}^2}, \\
S^{\text{hp}} &= \int_0^{t^*} \omega(s)^2 \frac{Y_{12}(s;1)Y_{12}(s;0)\{dN_1(s;1)+dN_1(s;0)\}}{\{Y_{12}(s;1)+Y_{12}(s;0)\}^2},
\end{align*}
respectively. Therefore, these hypotheses can be tested by comparing the standardized test statistics with the quantiles of the standard normal distribution.

\section{A comparison of the five strategies}\label{sec5}

There are two data structures: competing risks and semi-competing risks. In the former structure $\tilde{T}\wedge\tilde{R}$ is observed, whereas in the latter structure both $\tilde{T}$ and $\tilde{R}$ are observed. Table \ref{tab:compare} compares the five strategies in terms of data collection, hypothesis testing, and practical interpretation. Under the semi-competing risks data structure, all strategies can be applied. It is straightforward to transform a semi-competing risk structure into a competing risk structure by recording only the first event and disregarding subsequent observations. To be in line with the traditional competing risks analysis, we recompose $(W, \tilde{T} \wedge \tilde{R}, \Delta^T, \Delta^R)$ as $(W, \tilde{T} \wedge \tilde{R}, J)$, where $J = 0$ if $\Delta^T = \Delta^R = 0$ (only the censoring event is observed), $J = 1$ if $\tilde{T} \leq \tilde{R} $ and $\Delta^T = 1$ (only the primary outcome event is observed), and $J = 2$ if $\tilde{R}<\tilde{T}$ and $\Delta^R = 1$ (the intercurrent event is observed). The treatment policy strategy is not applicable if the intercurrent events prevent the measurement of primary outcome events, where only $\tilde{T} \wedge \tilde{R}$ instead of $(\tilde{T}, \tilde{R})$ is observed.

The five strategies differ in the scientific questions to answer and, therefore, incorporate intercurrent events differently. The treatment policy strategy considers the effect on primary outcome events with intercurrent events as natural. Because the intercurrent event is considered part of the treatment condition, any difference in the hazards of intercurrent events between the treatment groups reflects the effect of the treatment policy.
The composite variable strategy considers a composite of the primary outcome event and intercurrent event.
The while on treatment strategy estimand quantifies the total effect on the cumulative incidence of the primary outcome event, which is contributed by two sources: one is the effect of changing the hazard of primary outcome events, and the other is the effect of changing the hazard of intercurrent events. The first source is also quantified by the hypothetical strategy estimand $\tau^{\text{hp,I}}(t)$.
The principal stratum strategy considers a subpopulation defined by potential intercurrent events, typically where such events would not occur under either treatment condition.

\begin{table}
\centering
\caption{A comparison of five strategies} \label{tab:compare}
\begin{tabular}{p{0.16\textwidth}cccp{0.48\textwidth}}
\hline
\multirow{2}{*}{Strategy} & \multicolumn{2}{c}{Data collection} & Hypothesis & \multirow{2}{*}{Practical interpretation} \\
 & $\tilde{T}\wedge\tilde{R}$ & $(\tilde{T},\tilde{R})$ & testing & \\ \hline
Treatment policy  &   $\times$ & \checkmark & Log-rank test & Effect on primary outcome events with intercurrent events as natural\\
Composite variable  & \checkmark &  \checkmark & Log-rank test & Effect on the composite of primary outcome events and intercurrent events\\
While on treatment  & \checkmark &  \checkmark & Gray test & Effect on primary outcome events counted up to intercurrent events \\
Hypothetical (I, II) & \checkmark &  \checkmark & Log-rank test & Effect on primary outcome events adjusting for the hazard of intercurrent events  \\
Principal stratum & \checkmark &  \checkmark & No simple form & Effect on primary outcome events in a subpopulation determined by the potential occurrences of intercurrent events \\ \hline
\end{tabular}
\end{table}

In general, the hypothetical strategy and principal stratum strategy target some direct effects, while other strategies target some total effects. Identifying direct effects requires stronger assumptions, such as sequential ignorability or principal ignorability.
Under the multi-state model, let the original status, intercurrent events status, and primary outcome events status be three compartments. Then $\mu_w^{\text{cv}}(t)$ is the probability of being in either state of intercurrent events or primary outcome events under treatment condition $w$, whereas $\mu_w^{\text{wo}}(t)$ is the probability of being in the state of primary outcome events developed directly following the initial status \citep{buhler2023multistate}. If the data follow the semi-competing risks structure, $\mu_w^{\text{tp}}(t)$ represents the probability of being in the state of primary outcome events, regardless of whether intercurrent events occur before the primary outcome events.

We further illustrate the difference between these strategies with a simple example.

\begin{example}
Suppose that the hazard specific to the potential primary outcome event and intercurrent event are $\lambda_1(t;w) = a_w t$ and $\lambda_2(t;w) = c_w$, $w = 1, 0$, respectively. Therefore, the marginal distribution of the potential primary outcome event is Weibull, with $T(w) \sim Weibull(2, \sqrt{2/a_w})$. Suppose the intercurrent event neither prevents nor modifies the hazard of the primary outcome event. Table \ref{tab:eg} lists the analytical forms of $\mu_w^k(t)$. In the hypothetical scenario I, $c'_w = c_0$; in the hypothetical scenario II, $c'_w = 0$. In Supplementary Material B, we conduct a simulation study to assess the estimators and confidence intervals.
\end{example}

\begin{table}
\centering
\caption{Comparison of estimands under the five strategies in Example 1}\label{tab:eg}
\begin{tabular}{ll}
\hline
Strategy & $\mu_w^k (t)$, $0 \le t \le t^*$ \\ \hline
Treatment policy & $1 - e^{-a_w t^2/2}$ \\
Composite variable & $1 - e^{-a_w t^2/2 - c_w t}$ \\
While on treatment & $1-e^{-a_w t^2/2 - c_w t} - e^{c_w^2/2a_w} (2\pi c_w^2/a_w)^{1/2}\{\Phi(a_w^{1/2}(t+c_w/a_w)) - \Phi(c_w/a_w^{1/2})\}$ \\
Hypothetical & $1-e^{-a_w t^2/2 - c_w' t} - e^{c_w^{'2}/2a_w} (2\pi c_w^{'2}/a_w)^{1/2}\{\Phi(a_w^{1/2}(t+c_w'/a_w)) - \Phi(c_w'/a_w^{1/2})\}$ \\
Principal stratum & $\frac{1-e^{-a_w t^2/2 - c_w t} - e^{c_w^2/2a_w} (2\pi c_w^2/a_w)^{1/2}\{\Phi(a_w^{1/2}(t+c_w/a_w)) - \Phi(c_w/a_w^{1/2})\}}{1 - e^{c_w^2/2a_w} {(2\pi c_w^2/a_w)}^{1/2}\{\Phi(a_w^{1/2}(t^*+c_w/a_w)) - \Phi(c_w/a_w^{1/2})\}}$ \\
\hline
\end{tabular} \\
Notes: $\Phi(\cdot)$ denotes the cumulative distribution function of the standard normal distribution.
\end{table}

\section{Application to the LEADER Trial} \label{sec6}

In this section, we apply the five strategies with data from the LEADER Trial \citep{marso2016liraglutide} to evaluate the effect of liraglutide on cardiovascular outcomes. This trial was conducted at 410 clinical research sites in 32 countries as part of a large global phase 3a clinical development program. A sample including 9340 individuals with type 2 diabetes was randomly assigned to liraglutide (a glucagon-like peptide-1 receptor) or placebo for assessment of the long-term efficacy of liraglutide in preventing cardiovascular outcomes (including deaths from cardiovascular diseases, nonfatal myocardial infarction, or nonfatal stroke). However, a proportion of patients died due to other reasons before the measurement of primary outcome events. Specifically, 4668 individuals were allocated to liraglutide, among which 162 individuals died due to non-cardiovascular reasons; 4672 individuals were allocated to placebo, among which 169 individuals died due to non-cardiovascular reasons. The maximum follow-up time since treatment is 62.44 months, and the median is 46.09 months. The existence of non-cardiovascular death calls for caution when interpreting the primary outcome events using the intention-to-treat analysis because the occurrence of primary outcome events would not be well defined if they are truncated by death. 

As in the initial analysis of the LEADER Trial \citep{marso2016liraglutide}, we consider the occurrence of major adverse cardiovascular events (MACE, including non-fatal cardiovascular events and cardiovascular death) as the primary outcome event and non-cardiovascular death (NCVD) as the intercurrent event (Study I). Because the treatment policy is not applicable in this analysis as the primary outcome event may be prevented by the intercurrent event, we further conduct another analysis (Study II) where the primary outcome event is the all-cause death event and the intercurrent event is the occurrence of non-fatal major adverse cardiovascular events in Supplementary Material C. Considering ties, we record the event as the primary outcome event if the intercurrent event and primary outcome event occur at the same time. The 95\% confidence intervals of the cumulative incidences and time-varying treatment effects are calculated using the analytical formulas in Section \ref{sec3}. It is worth mentioning that these strategies result in different estimands and answer different scientific questions, some of which may not accord with the primary interest of regulatory agencies.

Of the individuals taking liraglutide, 608 had major adverse cardiovascular events and 137 died due to non-cardiovascular reasons; of those taking placebo, 694 had major adverse cardiovascular events and 133 died due to non-cardiovascular reasons. The composite variable strategy aims to evaluate the treatment effect on the cumulative incidence of the composite event, which comprises MACE and NCVD, thereby inducing a new endpoint. The while on treatment strategy aims to evaluate the treatment effect on the cumulative incidence of MACE, leaving the risk of NCVD as natural. The hypothetical strategy seeks to assess the treatment effect on the cumulative incidence of MACE by controlling the hazard of NCVD, either by maintaining a uniform hazard level or by completely eliminating the hazard. The principal stratum strategy aims to evaluate the treatment effect on the cumulative incidence of MACE in a subpopulation who would never experience NCVD regardless of treatment conditions.

\begin{figure}
\centering
\includegraphics[width=0.95\textwidth]{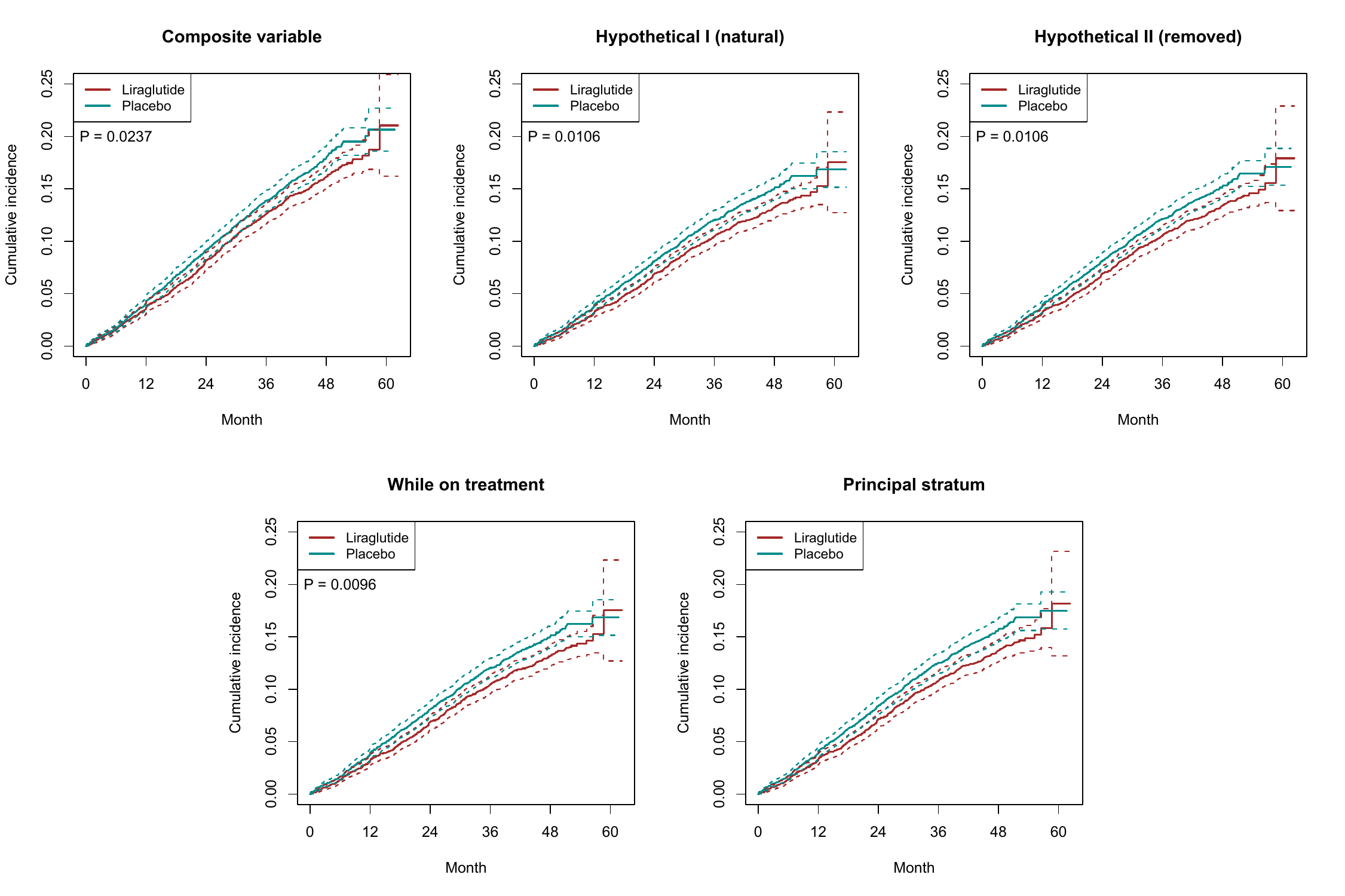}
\caption{Estimated cumulative incidences in Study I, with 95\% confidence intervals.} \label{fig1}
\end{figure}

\begin{figure}
\centering
\includegraphics[width=0.95\textwidth]{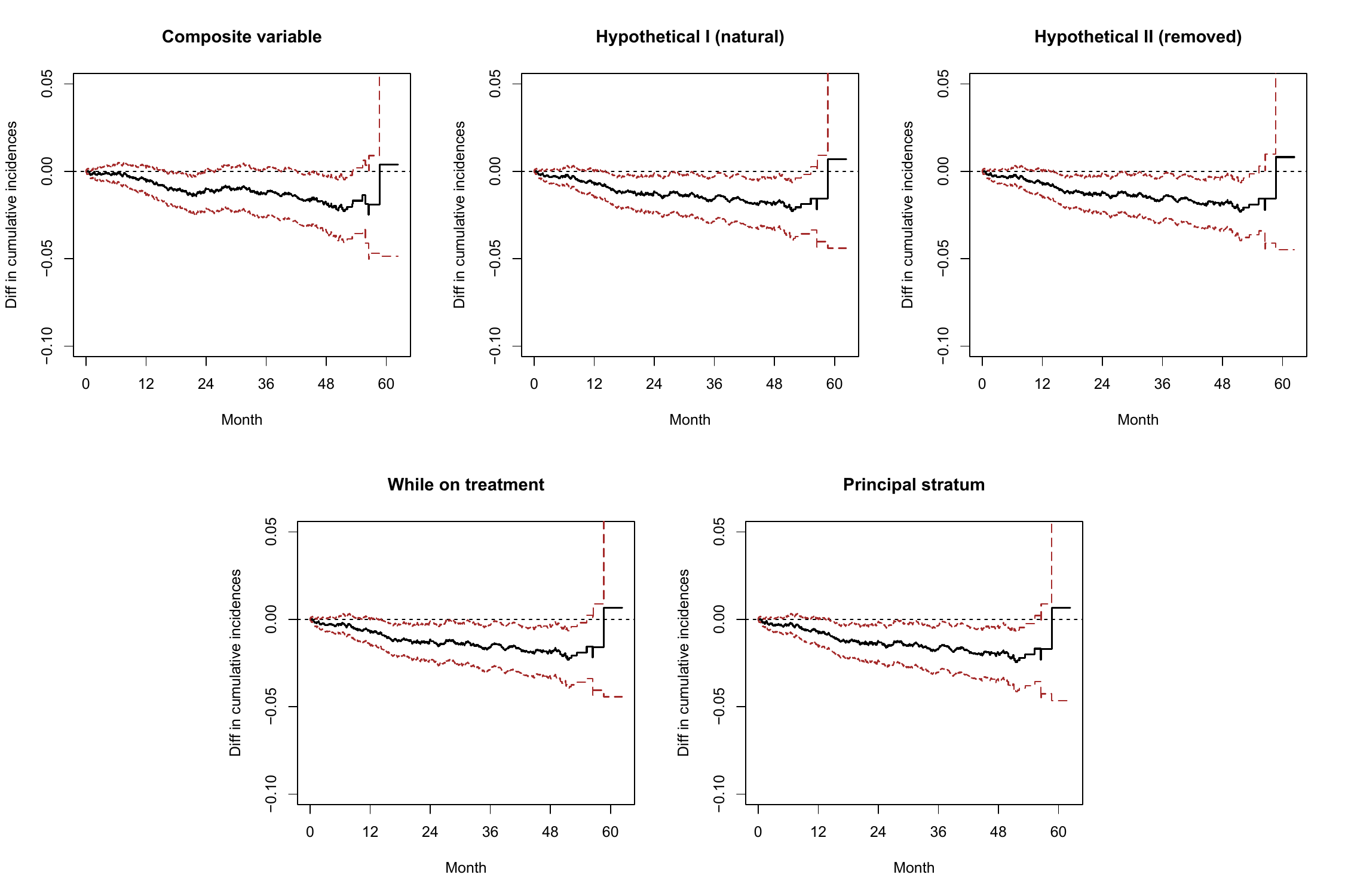}
\caption{Estimated treatment effects in Study I, with 95\% confidence intervals.} \label{fig2}
\end{figure}

Figure \ref{fig1} shows the estimated cumulative incidences, and Figure \ref{fig2} shows the estimated treatment effects under these four strategies. The estimated treatment effect shows a large variance due to the large censoring rate at the tail. Liraglutide can reduce the risk of MACE under all investigated strategies, with an increasing absolute risk difference over time. The $p$-value of the treatment effect in the composite variable strategy is 0.0237, indicating a significant effect of liraglutide on the combination of major adverse cardiovascular events and death. The $p$-value of the treatment effect in the hypothetical strategy is 0.0106, indicating a significant effect of liraglutide on the hazard of MACE. Due to the low hazard of death, these five estimates under comparison do not show a significant difference. By considering these strategies as sensitivity analyses to one another, the negligible difference strengthens the claim that liraglutide reduces the risk of MACE.


As expected, the cumulative incidences achieve the highest level under the composite variable strategy because the composite variable strategy considers the summarized occurrence of either the primary outcome event or the intercurrent event. Of note, the cumulative incidences under the principal stratum strategy are higher than those under the while on treatment strategy because the former incidences equal the latter divided by a factor of smaller than 1 (see Section \ref{estimation:ps}). The cumulative incidences in hypothetical scenario II are higher than those in hypothetical scenario I. Recall that in the former, we set the hazard of intercurrent events at zero, while in the latter, we only control the hazards of intercurrent events at an identical level under two treatment conditions (see Section \ref{estimation:hp}). The at-risk sets in the former would be larger, as some individuals who experienced intercurrent events could be maintained in the at-risk sets in the hypothetical scenario, resulting in higher cumulative incidences. The cumulative incidences under placebo in hypothetical scenario I are identical to those in the while on treatment strategy, confirming the earlier arguments in Section \ref{estimation:hp}.

\section{Discussion} \label{sec7}

The presence of intercurrent events poses challenges in the statistical analysis of clinical trials with time-to-event outcomes. In this paper, we construct the causal estimands and provide their estimators under the five strategies in ICH E9 (R1) addendum. The treatment effect is illustrated by a time-varying treatment effect curve, which contrasts the cumulative incidence functions associated with active treatment and placebo. From a causal inference perspective, these strategies address different scientific questions and may require different data to investigate them. As highlighted in the addendum, the choice of strategy depends on the specific scientific question that practitioners aim to answer. Some strategies may be challenging to interpret or guide practitioners in clinical practice. For example, the hypothetical strategy invisions a hypothetical scenario instead of directly exploring association using observed data, and the principal stratum strategy defines a causal effect in an unidentified target population. Therefore, drawing a meaningful causal conclusion requires discussion between statisticians and clinicians. From a regulatory perspective, the effectiveness of a drug can be reported based on estimated treatment effects at selected time points, along with standard errors and $p$-values, under a specific strategy. 

There are alternative causal estimands beyond the difference in cumulative incidences. Generally, estimands should be defined as a contrast of functionals of potential outcomes on a well-defined target population. Causal interpretations of such kinds of estimands may rely on specific contexts as well as some implications. For example, the average hazards ratio (AHR) integrates the ratio of hazards under the treated and under the control from treatment initiation to some time point $t$ \citep{kalbfleisch1981estimation, martinussen2020subtleties, prentice2022intention}, and the restricted mean time lost (RMTL) measures the survival time lost (or gained) by receiving active treatment compared to control in the time period from treatment initiation to some time point $t$ \citep{uno2015alternatives, zhao2016restricted}. Summarizing the treatment effect to a scalar provides convenience for inference and decision-making, although some information is lost. The five strategies in ICH E9 (R1) still apply to account for intercurrent events in these types of estimands.

In estimation, we focused on completely randomized trials where covariates are naturally balanced between treatment groups. Our proposed methods can be extended to incorporate covariates, allowing for the handling of stratified randomized trials or observational studies (real-world data) with an ignorable treatment assignment mechanism. With imbalanced baseline covariates, we can use semiparametric models, such as proportional hazards and additive hazards models, to estimate the hazards \citep{fine1999proportional, robins2000correcting, abbring2003identifiability, lu2008analysis, wang2017doubly, dukes2019doubly}. Alternatively, we might employ propensity-score-based methods, such as weighted Kaplan--Meier estimators, weighted Nelson--Aalen estimators, and propensity score matching \citep{xie2005adjusted, sugihara2010survival, austin2014use, mao2018propensity, hu2020modified}. G-formula and doubly robust estimation are also available considering propensity scores and failure time models simultaneously \citep{robins1986new, robins1997causal, zhang2012contrasting}. The variability of estimated nuisance models should be taken into consideration when investigating the asymptotics of estimators and constructing confidence intervals, which can pose a theoretical challenge. In Supplementary Material D, we provide semiparametric estimation approaches based on efficient influence functions that enjoy efficiency and robustness.

\section*{Acknowledgments}
Yuhao Deng and Shasha Han contributed equally.
The authors thank Kajsa Kvist and Henrik Ravn from Novo Nordisk A/S for comments.

\section*{Data availablility statement}
The \texttt{R} codes (version 4.4.1) for estimation and hypothesis testing are available on GitHub (\url{github.com/naiiife/ICHe9r1}).
The data that support the findings of this study are available from Novo Nordisk A/S. Restrictions apply to the availability of these data, which were used under license for this study.

\section*{Funding information}
This study was supported by grants from the National Key Research and Development Program of China (No.~2021YFF0901400), National Natural Science Foundation of China (No.~12026606, 12226005, 82304269), and the Chinese Academy of Medical Sciences (CAMS) Innovation Fund for Medical Sciences (No.~2023-I2M-3-008). This work was also supported by Novo Nordisk A/S.

\bibliographystyle{apalike}
\bibliography{ref}



\section*{Supplementary material}

# Supplementary Material for "Inference for Cumulative Incidences and Treatment Effects in Randomized Controlled Trials with Time-to-Event Outcomes under ICH E9 (R1)"

**Abstract**

This supplementary material includes four parts:

(A) Asymptotic variance in principal stratum strategy;

(B) Numerical studies;

(C) LEADER Trial: Study II;

(D) Estimation and inference in observational studies with ignorable treatment mechanism.

## A  Asymptotic variance in principal stratum strategy

Write

$$D_1 = \mu_w^{\mathrm{wo}}(t), \quad D_2 = 1 - \int_0^{t^*} e^{-\Lambda_{12}(s;w)} d\Lambda_2(s;w) = \exp\{-\Lambda_{12}(t^*;w)\} + \mu_w^{\mathrm{wo}}(t^*),$$

and the principal stratum estimand

$$\widehat{\mu}_w^{\mathrm{ps}}(t) = \frac{\widehat{\mu}_w^{\mathrm{wo}}(t)}{\exp\{-\widehat{\Lambda}_{12}(t^*;w)\} + \widehat{\mu}_w^{\mathrm{wo}}(t^*)} =: \frac{\widehat{D}_1}{\widehat{D}_2}.$$

The numerator

$$\begin{aligned}
\widehat{D}_1 &= \int_0^t \exp\{-\widehat{\Lambda}_{12}(s;w)\} d\widehat{\Lambda}_1(s,w) \\
&= \int_0^t \exp\{-\Lambda_{12}(s;s)\} \left\{1 - \int_0^s \frac{dM_1(u;w)}{Y_{12}(u;w)}\right\} \left\{d\Lambda_1(s,w) + \frac{dM_1(s;w)}{Y_{12}(s;w)}\right\} + o_p(n^{-1/2}) \\
&= \mu_w^{\mathrm{wo}}(t) + \int_0^t \{e^{-\Lambda_{12}(s;w)} + \mu_w^{\mathrm{wo}}(s) - \mu_w^{\mathrm{wo}}(t)\}\frac{dM_1(s;w)}{Y_{12}(s;w)} + \int_0^t \{\mu_w^{\mathrm{wo}}(s) - \mu_w^{\mathrm{wo}}(t)\}\frac{dM_2(s;w)}{Y_{12}(s;w)} + o_p(n^{-1/2}) \\
&= \mu_w^{\mathrm{wo}}(t) + \int_0^t A_1(s,t,w)\frac{dM_1(s;w)}{Y_{12}(s;w)} - \int_0^t B_1(s,t,w)\frac{dM_2(s;w)}{Y_{12}(s;w)} + o_p(n^{-1/2}).
\end{aligned}$$

1

Similarly, the denominator

$$\widehat{D}_2 = \exp\{-\Lambda_{12}(t^*; w)\} + \mu_w^{\text{wo}}(t^*) + \int_0^{t^*} A_2(s, t^*, w)\frac{dM_1(s;w)}{Y_{12}(s;w)} - \int_0^{t^*} B_2(s, t^*, w)\frac{dM_2(s;w)}{Y_{12}(s;w)} + o_p(n^{-1/2}).$$

By martingale theory,

$$\text{avar}(\widehat{D}_1) = \lim_{n\to\infty} \mathsf{E}\int_0^t A_1(s,t,w)^2 \frac{d\Lambda_1(s;w)}{Y_{12}(s;w)/n} + B_1(s,t,w)^2 \frac{d\Lambda_2(s;w)}{Y_{12}(s;w)/n}$$

$$= \int_0^{t^*} A_1(s,t,w)^2 \frac{d\Lambda_1(s;w)}{\Pr(\tilde{T}\wedge\tilde{R}\geq s, W=w)} + B_1(s,t,w)^2 \frac{d\Lambda_2(s;w)}{\Pr(\tilde{T}\wedge\tilde{R}\geq s, W=w)},$$

$$\text{avar}(\widehat{D}_2) = \lim_{n\to\infty} \mathsf{E}\int_0^{t^*} A_2(s,t^*,w)^2 \frac{d\Lambda_1(s;w)}{Y_{12}(s;w)/n} + B_2(s,t^*,w)^2 \frac{d\Lambda_2(s;w)}{Y_{12}(s;w)/n}$$

$$= \int_0^{t^*} A_2(s,t^*,w)^2 \frac{d\Lambda_1(s;w)}{\Pr(\tilde{T}\wedge\tilde{R}\geq s, W=w)} + B_2(s,t^*,w)^2 \frac{d\Lambda_2(s;w)}{\Pr(\tilde{T}\wedge\tilde{R}\geq s, W=w)},$$

$$\text{acov}(\widehat{D}_1, \widehat{D}_2) = \lim_{n\to\infty} \mathsf{E}\int_0^t A_1(s,t,w)A_2(s,t^*,w)\frac{d\Lambda_1(s;w)}{Y_{12}(s;w)/n} + B_1(s,t,w)B_2(s,t^*,w)\frac{d\Lambda_2(s;w)}{Y_{12}(s;w)/n}$$

$$= \int_0^{t^*} \frac{A_1(s,t,w)A_2(s,t^*,w)d\Lambda_1(s;w)}{\Pr(\tilde{T}\wedge\tilde{R}\geq s, W=w)} + \frac{B_1(s,t,w)B_2(s,t^*,w)d\Lambda_2(s;w)}{\Pr(\tilde{T}\wedge\tilde{R}\geq s, W=w)}.$$

So

$$\text{avar}\{\widehat{\mu}_w^{\text{ps}}(t)\} = \frac{1}{D_2^2}\text{avar}(\widehat{D}_1) + \frac{D_1^2}{D_2^4}\text{avar}(\widehat{D}_2) - 2\frac{D_1}{D_2^3}\text{acov}(\widehat{D}_1, \widehat{D}_2)$$

$$= \frac{1}{D_2^2}\left\{\text{avar}(\widehat{D}_1) + \mu_w^{\text{ps}}(t)^2\text{avar}(\widehat{D}_2) - 2\mu_w^{\text{ps}}(t)\text{acov}(\widehat{D}_1, \widehat{D}_2)\right\}$$

$$= \frac{1}{D_2^2}\int_0^{t^*}\{A_1(s,t,w) - \mu_w^{\text{ps}}(t)A_2(s,t^*,w)\}^2 \frac{d\Lambda_1(s;w)}{\Pr(\tilde{T}\wedge\tilde{R}\geq s, W=w)}$$

$$+ \frac{1}{D_2^2}\int_0^{t^*}\{B_1(s,t,w) - \mu_w^{\text{ps}}(t)B_2(s,t^*,w)\}^2 \frac{d\Lambda_2(s;w)}{\Pr(\tilde{T}\wedge\tilde{R}\geq s, W=w)}.$$

## B  Numerical studies

As in Example 1 of §5, suppose $a_1 = 0.2$, $a_0 = 0.1$, $c_1 = 0.3$, $c_0 = 0.2$. Each individual is assigned to active treatment or placebo with equal probability. The censoring time is assumed to be unifrom in $[4, 6]$ with probability 0.5 and be at 6 (end time of study) with probability 0.5. Suppose the sample size $n = 500$. We generate 10000 independent datasets and assess the performance of 95% asymptotic confidence intervals and bootstrap (with 100 resamplings) confidence intervals. Table S1 shows the coverage rate and Table S2 shows the average width of confidence intervals. Both types of confidence intervals have coverage rates close to the nominal level.

Table S1: Coverage rate of confidence intervals

| $t$ | 1 | 2 | 3 | 4 | 5 | 6 |
|---|---|---|---|---|---|---|
| Confidence interval by analytical form | | | | | | |
| Treatment policy | 0.9518 | 0.9511 | 0.9487 | 0.9502 | 0.9501 | 0.9514 |
| Composite variable | 0.9451 | 0.9524 | 0.9477 | 0.9491 | 0.9466 | 0.9426 |
| While on treatment | 0.9510 | 0.9506 | 0.9508 | 0.9501 | 0.9513 | 0.9518 |
| Hypothetical I (natural) | 0.9511 | 0.9529 | 0.9529 | 0.9552 | 0.9625 | 0.9656 |
| Hypothetical II (removed) | 0.9495 | 0.9510 | 0.9506 | 0.9485 | 0.9472 | 0.9476 |
| Principal stratum | 0.9487 | 0.9498 | 0.9493 | 0.9533 | 0.9513 | 0.9657 |
| Confidence interval by bootstrap | | | | | | |
| Treatment policy | 0.9472 | 0.9478 | 0.9453 | 0.9469 | 0.9444 | 0.9456 |
| Composite variable | 0.9427 | 0.9498 | 0.9440 | 0.9457 | 0.9400 | 0.9307 |
| While on treatment | 0.9434 | 0.9426 | 0.9421 | 0.9398 | 0.9403 | 0.9365 |
| Hypothetical I (natural) | 0.9431 | 0.9448 | 0.9425 | 0.9401 | 0.9343 | 0.9212 |
| Hypothetical II (removed) | 0.9446 | 0.9449 | 0.9478 | 0.9411 | 0.9386 | 0.9279 |
| Principal stratum | 0.9453 | 0.9434 | 0.9407 | 0.9408 | 0.9338 | 0.9346 |

Table S2: Average width of confidence intervals

| $t$ | 1 | 2 | 3 | 4 | 5 | 6 |
|---|---|---|---|---|---|---|
| Confidence interval by analytical form | | | | | | |
| Treatment policy | 0.0900 | 0.1505 | 0.1702 | 0.1583 | 0.1369 | 0.1185 |
| Composite variable | 0.1553 | 0.1716 | 0.1496 | 0.1155 | 0.0866 | 0.0683 |
| While on treatment | 0.0831 | 0.1349 | 0.1602 | 0.1706 | 0.1764 | 0.1804 |
| Hypothetical I (natural) | 0.0863 | 0.1404 | 0.1573 | 0.1487 | 0.1353 | 0.1273 |
| Hypothetical II (removed) | 0.0985 | 0.1808 | 0.2251 | 0.2298 | 0.2184 | 0.2131 |
| Principal stratum | 0.1676 | 0.2419 | 0.2385 | 0.2003 | 0.1634 | 0.1438 |
| Confidence interval by bootstrap | | | | | | |
| Treatment policy | 0.0892 | 0.1492 | 0.1686 | 0.1569 | 0.1356 | 0.1165 |
| Composite variable | 0.1539 | 0.1701 | 0.1483 | 0.1147 | 0.0856 | 0.0653 |
| While on treatment | 0.0818 | 0.1321 | 0.1557 | 0.1646 | 0.1684 | 0.1699 |
| Hypothetical I (natural) | 0.0849 | 0.1374 | 0.1519 | 0.1402 | 0.1209 | 0.1083 |
| Hypothetical II (removed) | 0.0976 | 0.1791 | 0.2220 | 0.2260 | 0.2120 | 0.1972 |
| Principal stratum | 0.1663 | 0.2381 | 0.2317 | 0.1918 | 0.1533 | 0.1287 |

# C  LEADER Trial: Study II

In Study II, the primary outcome event is all-cause death and the intercurrent event is the occurrence of non-fatal major adverse cardiovascular events (MACE). Of the individuals taking liraglutide, 381 individuals died and 364 experienced non-fatal major adverse cardiovascular events; and of those taking placebo, 447 died and 380 experienced non-fatal major adverse cardiovascular events. Treatment policy strategy aims to evaluate the treatment effect on the incidence of death (no matter with MACE or not) leaving the risk of MACE as natural. Composite variable strategy aims to evaluate the treatment effect on the incidence of the composite event comprised of death and MACE (inducing a new endpoint). While on treatment strategy aims to evaluate the treatment effect on the incidence of non-MACE death (non-cardiovascular death) leaving the risk of MACE as natural. Hypothetical strategy aims to evaluate the treatment effect on the incidence of death by controlling the hazard of MACE. Principal stratum strategy aims to evaluate the treatment effect on the incidence of death in a subpopulation who would never experience MACE regardless of treatment conditions.

Figure S1 shows the estimated cumulative incidences and Figure S2 shows the estimated treatment effects under these five strategies. A reduced risk of death is found for liraglutide under all the investigated strategies. Of note, the cumulative incidences under the treatment policy strategy are higher than those under the while on treatment strategy, because the treatment policy strategy has counted more primary outcome events than the while on treatment strategy. The $p$-value of the treatment effect in the treatment policy strategy is 0.0170, indicating a significant effect of liraglutide on all-cause death with non-fatal cardiovascular events as intercurrent events. The $p$-value of the treatment effect in the composite variable strategy is 0.0237, identical to the $p$-value in the composite variable strategy of Study I because the composites of primary outcomes and intercurrent events are exactly identical in these two analyses. The $p$-value of the treatment effect in the hypothetical strategy is 0.0148, indicating a significant effect of liraglutide on the hazard of all-cause death among the at-risk individuals without non-fatal cardiovascular events.

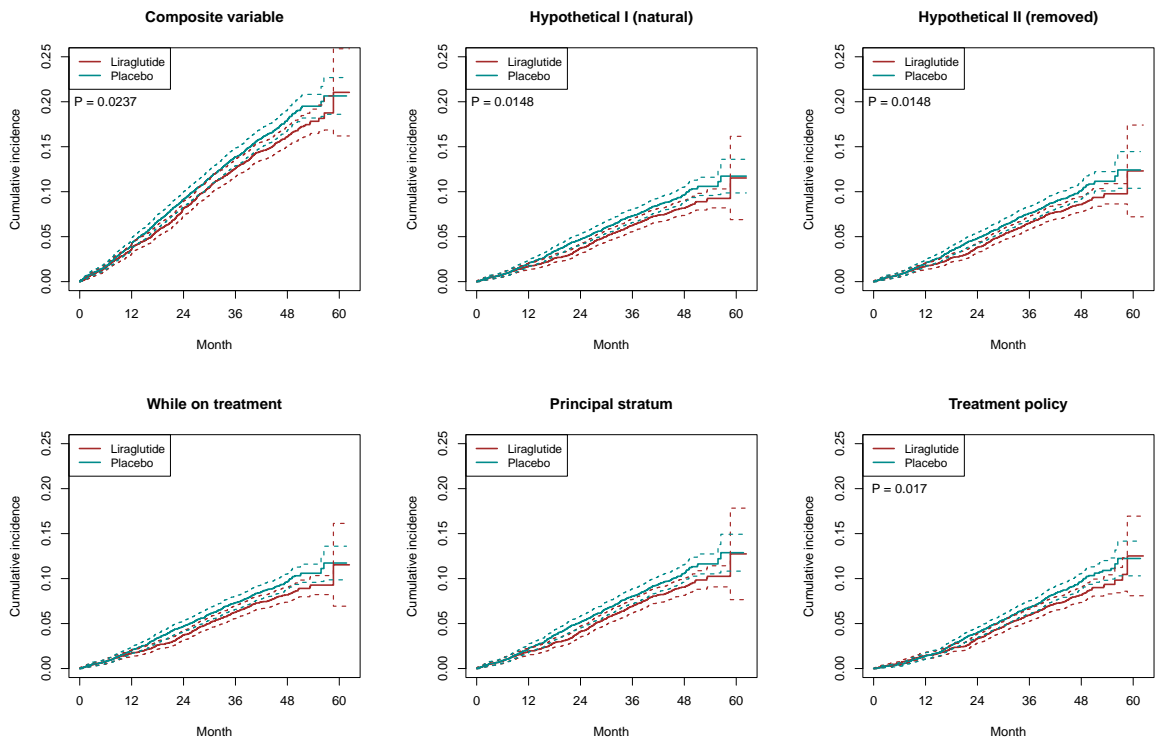

Figure S1: Estimated cumulative incidences in Study II, with 95% confidence intervals.

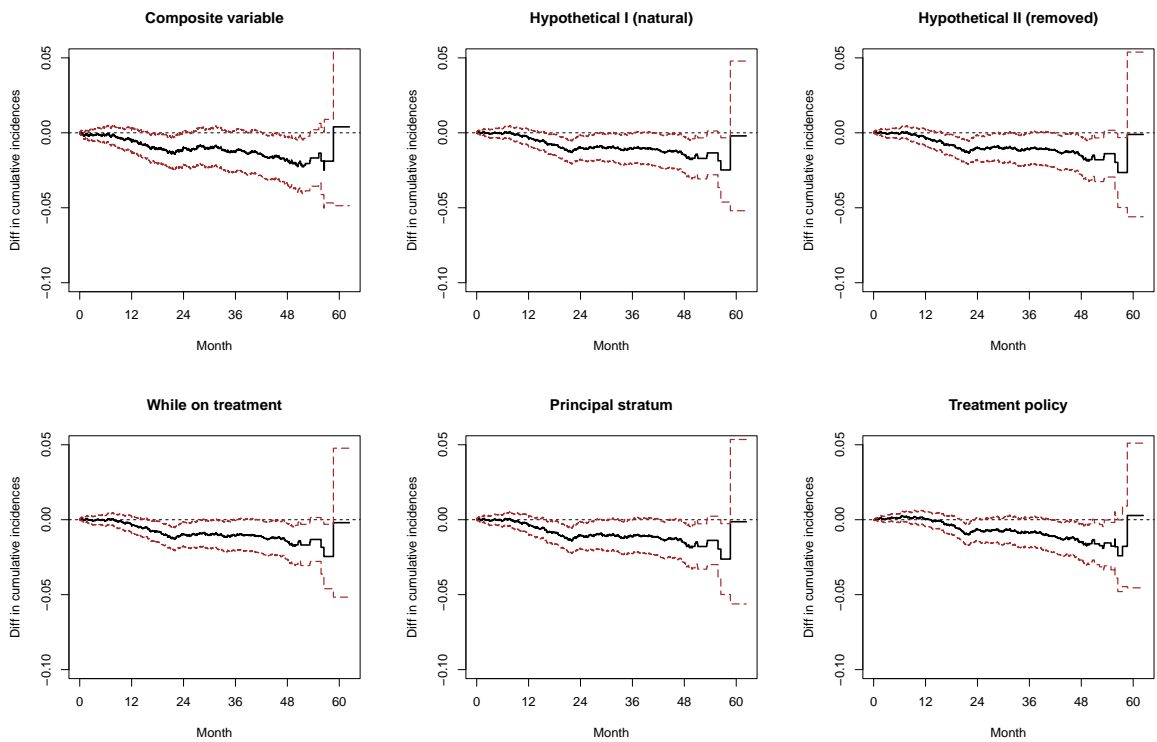

Figure S2: Estimated treatment effects in Study II, with 95% confidence intervals.

5

# D   Estimation and inference in observational studies with ignorable treatment mechanism

Suppose that some baseline covariates $X$ are collected in an observational study, and Assumptions 1–4 are replaced with conditional versions (conditional randomization, conditional random censoring, conditional positivity with positive treatment probability, conditional principal ignorability). The estimands under five strategies are not affected as we are still interested in the population level cumulative incidences and treatment effect. Identification is straightforward by first conditioning on covariates and then taking average over the population (except for the principal stratum strategy where the target population is not the whole population). For estimation, a basic idea is to use generalized (weighted) Nelson-Aalen estimators by inverse probability weighting (IPW), which only requires estimation of the propensity score. In stratified randomized trials, the propensity score may be known by design. Alternatively, we can use functional estimators like Cox regression to model the cause-specific hazards. However, these two types of estimators are vulnerable to model misspecification.

In the context of having baseline covariates, we present semiparametric estimators through five strategies that rely on efficient influence functions [Kosorok (2008); Kennedy (2022)]. We posit that all models belong to the Donsker class (for example, we can use Cox proportional hazards models for cause-specific hazards). The efficient inflence functions in competing risks have been studied in literature [Martinussen and Stensrud (2023)]. The semiparametric estimators have robustness to model misspecification compared with inverse probability weighting estimators or functional estimators. When all models are correctly specified, the provided estimators have smaller variance and hence can increase power. We use $\mathsf{E}_n(\cdot)$ to denote the empirical average over the sample.

## D.1   Treatment policy strategy

**Models**

1. Propensity score model $\Pr(W = w \mid X = x)$, estimated by $\widehat{\Pr}(W = w \mid X = x)$.

2. Hazard model for the primary outcome event $\Lambda(t|w, x)$, estimated by $\widehat{\Lambda}(t|w, x)$.

3. Hazard model for censoring $\Lambda_C(t|w, x)$, estimated by $\widehat{\Lambda}_C(t|w, x)$.

Denote $M(t|w, x) = \int_0^t dN(s; w) - Y(s; w) d\widehat{\Lambda}(s|w, x)$.

**Efficient influence functions**

$$\mathsf{IF}\{\mu_w^{\mathrm{tp}}(t)\} = \frac{\mathsf{I}\{W=w\}}{\mathsf{Pr}(W=w\mid X)} e^{-\Lambda(t|w,X)} \int_0^t \frac{dM(s|w,X)}{e^{-\Lambda(s|w,X)-\Lambda_C(s|w,X)}} - e^{-\Lambda(t|w,X)} + 1 - \mu_w^{\mathrm{tp}}(t),$$

$$\mathsf{IF}\{\tau^{\mathrm{tp}}(t)\} = \mathsf{IF}\{\mu_1^{\mathrm{tp}}(t)\} - \mathsf{IF}\{\mu_0^{\mathrm{tp}}(t)\}.$$

**Estimators**

$$\widehat{\mu}_w^{\mathrm{tp}}(t) = \mathsf{E}_n\left[\frac{\mathsf{I}\{W=w\}}{\widehat{\mathsf{Pr}}(W=w\mid X)} e^{-\widehat{\Lambda}(t|w,X)} \int_0^t \frac{dM(s|w,X)}{e^{-\widehat{\Lambda}(s|w,X)-\widehat{\Lambda}_C(s|w,X)}} + 1 - e^{-\widehat{\Lambda}(t|w,X)}\right],$$

$$\widehat{\tau}^{\mathrm{tp}}(t) = \widehat{\mu}_1^{\mathrm{tp}}(t) - \widehat{\mu}_0^{\mathrm{tp}}(t).$$

**Asymptotics**

$$n^{1/2}\{\widehat{\mu}_w^{\mathrm{tp}}(t) - \mu_w^{\mathrm{tp}}(t)\} \to_d N(0, \mathsf{E}[\mathsf{IF}\{\mu_w^{\mathrm{tp}}(t)\}]^2),$$

$$n^{1/2}\{\widehat{\tau}^{\mathrm{tp}}(t) - \tau^{\mathrm{tp}}(t)\} \to_d N(0, \mathsf{E}[\mathsf{IF}\{\tau^{\mathrm{tp}}(t)\}]^2).$$

**Robustness**  The estimator is consistent if (1) Model $\{1,3\}$ or (2) Model $\{2\}$ is correctly specified.

### D.2  Composite variable strategy

**Models**

1. Propensity score model $\mathsf{Pr}(W=w\mid X=x)$, estimated by $\widehat{\mathsf{Pr}}(W=w\mid X=x)$.

2. Hazard model for the combined outcome event $\Lambda_{12}(t|w,x)$, estimated by $\widehat{\Lambda}_{12}(t|w,x)$.

3. Hazard model for censoring $\Lambda_C(t|w,x)$, estimated by $\widehat{\Lambda}_C(t|w,x)$.

Denote $M_{12}(t|w,x) = \int_0^t dN_{12}(s;w) - Y_{12}(s;w)d\widehat{\Lambda}_{12}(s|w,x)$.

**Efficient influence functions**

$$\mathsf{IF}\{\mu_w^{\mathrm{cv}}(t)\} = \frac{\mathsf{I}\{W=w\}}{\mathsf{Pr}(W=w\mid X)} e^{-\Lambda_{12}(t|w,X)} \int_0^t \frac{dM_{12}(s|w,X)}{e^{-\Lambda_{12}(s|w,X)-\Lambda_C(s|w,X)}} - e^{-\Lambda_{12}(t|w,X)} + 1 - \mu_w^{\mathrm{cv}}(t),$$

$$\mathsf{IF}\{\tau^{\mathrm{cv}}(t)\} = \mathsf{IF}\{\mu_1^{\mathrm{cv}}(t)\} - \mathsf{IF}\{\mu_0^{\mathrm{cv}}(t)\}.$$

**Estimators**

$$\widehat{\mu}_w^{\text{cv}}(t) = \mathsf{E}_n\left[\frac{\mathsf{I}\{W=w\}}{\widehat{\Pr}(W=w\mid X)}e^{-\widehat{\Lambda}_{12}(t|w,X)}\int_0^t \frac{dM_{12}(s|w,X)}{e^{-\widehat{\Lambda}_{12}(s|w,X)-\widehat{\Lambda}_C(s|w,X)}} + 1 - e^{-\widehat{\Lambda}_{12}(t|w,X)}\right],$$

$$\widehat{\tau}^{\text{cv}}(t) = \widehat{\mu}_1^{\text{cv}}(t) - \widehat{\mu}_0^{\text{cv}}(t).$$

**Asymptotics**

$$n^{1/2}\{\widehat{\mu}_w^{\text{cv}}(t) - \mu_w^{\text{cv}}(t)\} \to_d N(0, \mathsf{E}[\mathsf{IF}\{\mu_w^{\text{cv}}(t)\}]^2),$$

$$n^{1/2}\{\widehat{\tau}^{\text{cv}}(t) - \tau^{\text{cv}}(t)\} \to_d N(0, \mathsf{E}[\mathsf{IF}\{\tau^{\text{cv}}(t)\}]^2).$$

**Robustness** The estimator is consistent if (1) Model {1,3} or (2) Model {2} is correctly specified.

### D.3 While on treatment strategy

**Models**

1. Propensity score model $\Pr(W=w\mid X=x)$, estimated by $\widehat{\Pr}(W=w\mid X=x)$.

2. Hazard model for the primary outcome event $\Lambda_1(t|w,x)$, estimated by $\widehat{\Lambda}_1(t|w,x)$.

3. Hazard model for the intercurrent event $\Lambda_2(t|w,x)$, estimated by $\widehat{\Lambda}_2(t|w,x)$.

4. Hazard model for censoring $\Lambda_C(t|w,x)$, estimated by $\widehat{\Lambda}_C(t|w,x)$.

Denote $M_j(t|w,x) = \int_0^t dN_j(s;w) - Y_j(s;w)d\widehat{\Lambda}_j(s|w,x)$, $j=1,2$, $\Lambda_{12}(t|w,x) = \Lambda_1(t|w,x) + \Lambda_2(t|w,x)$ and $\mu^{\text{wo}}(t|w,x) = \int_0^t e^{-\Lambda_{12}(s|w,x)}d\Lambda_1(s|w,x)$.

**Efficient influence functions**

$$\mathsf{IF}\{\mu_w^{\text{wo}}(t)\} = \frac{\mathsf{I}\{W=w\}}{\Pr(W=w\mid X)}\int_0^t\left\{1 - \frac{\mu^{\text{wo}}(t|w,X) - \mu^{\text{wo}}(s|w,X)}{e^{-\Lambda_{12}(s|w,X)}}\right\}\frac{dM_1(s|w,X)}{e^{-\Lambda_C(s|w,X)}}$$
$$- \frac{\mathsf{I}\{W=w\}}{\Pr(W=w\mid X)}\int_0^t \frac{\mu^{\text{wo}}(t|w,X) - \mu^{\text{wo}}(s|w,X)}{e^{-\Lambda_{12}(s|w,X)}}\frac{dM_2(s|w,X)}{e^{-\Lambda_C(s|w,X)}} + \mu^{\text{wo}}(t|w,X) - \mu_w^{\text{wo}}(t),$$

$$\mathsf{IF}\{\tau^{\text{wo}}(t)\} = \mathsf{IF}\{\mu_1^{\text{wo}}(t)\} - \mathsf{IF}\{\mu_0^{\text{wo}}(t)\}.$$

**Estimators**

$$\widehat{\mu}_w^{\text{wo}}(t) = \mathsf{E}_n \left[ \frac{\mathsf{I}\{W=w\}}{\widehat{\mathsf{Pr}}(W=w \mid X)} \int_0^t \left\{ 1 - \frac{\widehat{\mu}^{\text{wo}}(t|w,X) - \widehat{\mu}^{\text{wo}}(s|w,X)}{e^{-\widehat{\Lambda}_{12}(s|w,X)}} \right\} \frac{dM_1(s|w,X)}{e^{-\widehat{\Lambda}_C(s|w,X)}} \right.$$
$$\left. - \frac{\mathsf{I}\{W=w\}}{\widehat{\mathsf{Pr}}(W=w \mid X)} \int_0^t \frac{\widehat{\mu}^{\text{wo}}(t|w,X) - \widehat{\mu}^{\text{wo}}(s|w,X)}{e^{-\widehat{\Lambda}_{12}(s|w,X)}} \frac{dM_2(s|w,X)}{e^{-\widehat{\Lambda}_C(s|w,X)}} + \widehat{\mu}^{\text{wo}}(t|w,X) \right],$$
$$\widehat{\tau}^{\text{wo}}(t) = \widehat{\mu}_1^{\text{wo}}(t) - \widehat{\mu}_0^{\text{wo}}(t).$$

**Asymptotics**

$$n^{1/2}\{\widehat{\mu}_w^{\text{wo}}(t) - \mu_w^{\text{wo}}(t)\} \to_d N(0, \mathsf{E}[\mathsf{IF}\{\mu_w^{\text{wo}}(t)\}]^2),$$
$$n^{1/2}\{\widehat{\tau}^{\text{wo}}(t) - \tau^{\text{wo}}(t)\} \to_d N(0, \mathsf{E}[\mathsf{IF}\{\tau^{\text{wo}}(t)\}]^2).$$

**Robustness** The estimator is consistent if (1) Model $\{2,3\}$ or (2) $\{1,2,4\}$ or (3) $\{1,3,4\}$ is correctly specified.

## D.4 Hypothetical strategy

**Models**

1. Propensity score model $\mathsf{Pr}(W = w \mid X = x)$, estimated by $\widehat{\mathsf{Pr}}(W = w \mid X = x)$.

2. Hazard model for the primary outcome event $\Lambda_1(t|w,x)$, estimated by $\widehat{\Lambda}_1(t|w,x)$.

3. Hazard model for the intercurrent event $\Lambda_2(t|w,x)$, estimated by $\widehat{\Lambda}_2(t|w,x)$.

4. Hazard model for censoring $\Lambda_C(t|w,x)$, estimated by $\widehat{\Lambda}_C(t|w,x)$.

Denote $M_j(t|w,x) = \int_0^t dN_j(s;w) - Y_j(s;w)d\widehat{\Lambda}_j(s|w,x)$, $j = 1, 2$, $\Lambda_{12}(t|w,x) = \Lambda_1(t|w,x) + \Lambda_2(t|w,x)$ and $\mu^{\text{hp,I}}(t|w,x) = \int_0^t e^{-\Lambda_1(s|w,x) - \Lambda_2(s|0,x)} d\Lambda_1(s|w,x)$.

**Efficient influence functions** Scenario I: controlling the hazard of the intercurrent event.

$$\mathsf{IF}\{\mu_w^{\text{hp,I}}(t)\} = \frac{\mathsf{I}\{W=w\}}{\mathsf{Pr}(W=w \mid X)} \int_0^t \left\{ \frac{e^{-\Lambda_2(s|0,X)}}{e^{-\Lambda_2(s|w,X)}} - \frac{\mu^{\text{hp,I}}(t|w,X) - \mu^{\text{hp,I}}(s|w,X)}{e^{-\Lambda_{12}(s|w,X)}} \right\} \frac{dM_1(s|w,X)}{e^{-\Lambda_C(s|w,X)}}$$
$$- \frac{\mathsf{I}\{W=0\}}{\mathsf{Pr}(W=0 \mid X)} \int_0^t \frac{\mu^{\text{hp,I}}(t|w,X) - \mu^{\text{hp,I}}(s|w,X)}{e^{-\Lambda_{12}(s|0,X)}} \frac{dM_2(s|0,X)}{e^{-\Lambda_C(s|0,X)}} + \mu^{\text{hp,I}}(t|w,X) - \mu_w^{\text{hp,I}}(t),$$
$$\mathsf{IF}\{\tau^{\text{hp,I}}(t)\} = \mathsf{IF}\{\mu_1^{\text{hp,I}}(t)\} - \mathsf{IF}\{\mu_0^{\text{hp,I}}(t)\}.$$

9

Scenario II: removing the hazard of the intercurrent event.

$$\mathsf{IF}\{\mu_w^{\mathrm{hp,II}}(t)\} = \frac{\mathsf{I}\{W=w\}}{\mathsf{Pr}(W=w\mid X)} e^{-\Lambda_1(t|w,X)} \int_0^t \frac{dM_1(s|w,X)}{e^{-\Lambda_{12}(s|w,X)-\Lambda_C(s|w,X)}} - e^{-\Lambda_1(t|w,X)} + 1 - \mu_w^{\mathrm{hp,II}}(t),$$

$$\mathsf{IF}\{\tau^{\mathrm{hp,II}}(t)\} = \mathsf{IF}\{\mu_1^{\mathrm{hp,II}}(t)\} - \mathsf{IF}\{\mu_0^{\mathrm{hp,II}}(t)\}.$$

**Estimators**   Scenario I: controlling the hazard of the intercurrent event.

$$\widehat{\mu}_w^{\mathrm{hp,I}}(t) = \mathsf{E}_n\Bigg[\frac{\mathsf{I}\{W=w\}}{\widehat{\mathsf{Pr}}(W=w\mid X)}\int_0^t\left\{\frac{e^{-\widehat{\Lambda}_2(s|0,X)}}{e^{-\widehat{\Lambda}_2(s|w,X)}} - \frac{\widehat{\mu}^{\mathrm{hp,I}}(t|w,X) - \widehat{\mu}^{\mathrm{hp,I}}(s|w,X)}{e^{-\widehat{\Lambda}_{12}(s|w,X)}}\right\}\frac{dM_1(s|w,X)}{e^{-\widehat{\Lambda}_C(s|w,X)}}$$
$$-\frac{\mathsf{I}\{W=0\}}{\widehat{\mathsf{Pr}}(W=0\mid X)}\int_0^t \frac{\widehat{\mu}^{\mathrm{hp,I}}(t|w,X) - \widehat{\mu}^{\mathrm{hp,I}}(s|w,X)}{e^{-\widehat{\Lambda}_{12}(s|0,X)}}\frac{dM_2(s|0,X)}{e^{-\widehat{\Lambda}_C(s|0,X)}} + \widehat{\mu}^{\mathrm{hp,I}}(t|w,X)\Bigg],$$

$$\widehat{\tau}^{\mathrm{hp,I}}(t) = \widehat{\mu}_1^{\mathrm{hp,I}}(t) - \widehat{\mu}_0^{\mathrm{hp,I}}(t).$$

Scenario II: removing the hazard of the intercurrent event.

$$\widehat{\mu}_w^{\mathrm{hp,II}}(t) = \mathsf{E}_n\Bigg[\frac{\mathsf{I}\{W=w\}}{\widehat{\mathsf{Pr}}(W=w\mid X)} e^{-\widehat{\Lambda}_1(t|w,X)} \int_0^t \frac{dM_1(s|w,X)}{e^{-\widehat{\Lambda}_{12}(s|w,X)-\widehat{\Lambda}_C(s|w,X)}} + 1 - e^{-\widehat{\Lambda}_1(t|w,X)}\Bigg].$$

$$\widehat{\tau}^{\mathrm{hp,II}}(t) = \widehat{\mu}_1^{\mathrm{hp,II}}(t) - \widehat{\mu}_0^{\mathrm{hp,II}}(t).$$

**Asymptotics**   In both scenarios ($j \in \{\mathrm{I}, \mathrm{II}\}$),

$$n^{1/2}\{\widehat{\mu}_w^{\mathrm{hp},j}(t) - \mu_w^{\mathrm{hp},j}(t)\} \to_d N(0, \mathsf{E}[\mathsf{IF}\{\mu_w^{\mathrm{hp},j}(t)\}]^2),$$

$$n^{1/2}\{\widehat{\tau}^{\mathrm{hp},j}(t) - \tau^{\mathrm{hp},j}(t)\} \to_d N(0, \mathsf{E}[\mathsf{IF}\{\tau^{\mathrm{hp},j}(t)\}]^2).$$

**Robustness**   The estimator is consistent if (1) Model $\{2,3\}$ or (2) $\{1,2,4\}$ or (3) $\{1,3,4\}$ is correctly specified.

### D.5   Principal stratum strategy

Typically, identification of the principal strtaum estimand requires an additional monotonicity, otherwise the proportion of the stratum $\{R(1) = R(0) = \infty\}$ is not identifiable. To reduce the assumptions involved and computational complexity, we consider a pseudo estimand $\mathsf{Pr}(T(1) < t \mid R(1) = \infty) - \mathsf{Pr}(T(0) < t \mid R(0) = \infty)$ instead. This estimand may not have a meaningful causal interpretation unless there is no systematic difference between the populations of $\{R(1) = \infty\}$ and $\{R(0) = \infty\}$.

**Models**

1. Propensity score model $\Pr(W = w \mid X = x)$, estimated by $\widehat{\Pr}(W = w \mid X = x)$.

2. Hazard model for the primary outcome event $\Lambda_1(t|w,x)$, estimated by $\widehat{\Lambda}_1(t|w,x)$.

3. Hazard model for the intercurrent event $\Lambda_2(t|w,x)$, estimated by $\widehat{\Lambda}_2(t|w,x)$.

4. Hazard model for censoring $\Lambda_C(t|w,x)$, estimated by $\widehat{\Lambda}_C(t|w,x)$.

Denote $M_j(t|w,x) = \int_0^t dN_j(s;w) - Y_j(s;w)d\widehat{\Lambda}_j(s|w,x)$, $j=1,2$, $\Lambda_{12}(t|w,x) = \Lambda_1(t|w,x) + \Lambda_2(t|w,x)$, $\mu^{\mathrm{wo}}(t|w,x) = \int_0^t e^{-\Lambda_{12}(s|w,x)} d\Lambda_1(s|w,x)$ and $\mu^{\mathrm{cv}}(t|w,x) = \int_0^t e^{-\Lambda_{12}(s|w,x)} d\Lambda_{12}(s|w,x)$.

**Efficient influence functions**

$$\mathsf{IF}\{\mu_w^{\mathrm{ps}}(t)\} = \frac{\mathsf{IF}\{\mu_w^{\mathrm{wo}}(t)\}}{1 - \mu_w^{\mathrm{cv}}(t^*) + \mu_w^{\mathrm{wo}}(t^*)} + \frac{\mu_w^{\mathrm{wo}}(t)[\mathsf{IF}\{\mu_w^{\mathrm{cv}}(t^*) - \mu_w^{\mathrm{wo}}(t^*)\}]}{\{1 - \mu_w^{\mathrm{cv}}(t^*) + \mu_w^{\mathrm{wo}}(t^*)\}^2},$$
$$\mathsf{IF}\{\tau^{\mathrm{ps}}(t)\} = \mathsf{IF}\{\mu_1^{\mathrm{ps}}(t)\} - \mathsf{IF}\{\mu_0^{\mathrm{ps}}(t)\}.$$

**Estimators**

$$\widehat{\mu}_w^{\mathrm{wo}}(t) = \frac{\widehat{\mathsf{IF}}\{\widehat{\mu}_w^{\mathrm{wo}}(t)\} + \widehat{\mu}_w^{\mathrm{wo}}(t)}{1 - \widehat{\mu}_w^{\mathrm{cv}}(t^*) + \widehat{\mu}_w^{\mathrm{wo}}(t^*)} + \frac{\widehat{\mu}_w^{\mathrm{wo}}(t)[\widehat{\mathsf{IF}}\{\widehat{\mu}_w^{\mathrm{cv}}(t^*) - \widehat{\mu}_w^{\mathrm{wo}}(t^*)\}]}{\{1 - \widehat{\mu}_w^{\mathrm{cv}}(t^*) + \widehat{\mu}_w^{\mathrm{wo}}(t^*)\}^2},$$
$$\widehat{\tau}^{\mathrm{ps}}(t) = \widehat{\mu}_1^{\mathrm{ps}}(t) - \widehat{\mu}_0^{\mathrm{ps}}(t).$$

**Asymptotics**

$$n^{1/2}\{\widehat{\mu}_w^{\mathrm{ps}}(t) - \mu_w^{\mathrm{ps}}(t)\} \to_d N(0, \mathsf{E}[\mathsf{IF}\{\mu_w^{\mathrm{ps}}(t)\}]^2),$$
$$n^{1/2}\{\widehat{\tau}^{\mathrm{ps}}(t) - \tau^{\mathrm{ps}}(t)\} \to_d N(0, \mathsf{E}[\mathsf{IF}\{\tau^{\mathrm{ps}}(t)\}]^2).$$

**Robustness** The estimator is consistent if (1) Model $\{2,3\}$ or (2) $\{1,2,4\}$ or (3) $\{1,3,4\}$ is correctly specified.



\end{document}